\documentclass[preprint,14pt,compress]{elsarticle}

\usepackage{amsfonts,amssymb,amsmath}
\usepackage{indentfirst}
\usepackage{amsthm}
\usepackage{graphicx}
\usepackage{subfig}
\usepackage{epstopdf}
\usepackage{float}
\usepackage{mathrsfs}
\usepackage{multirow}
\usepackage{array}
\usepackage{bm}
\usepackage{color}
\usepackage{ulem}
\emergencystretch=\maxdimen
\journal{arXiv}

\begin{document}

\begin{frontmatter}

\title{Numerical simulation of a coupled system of Maxwell equations and a gas dynamic model}

\author[CQU]{Maohui Lyu}
\ead{marvin0639@gmail.com}
\author[PDU]{Weng Cho Chew}
\ead{wcchew@purdue.edu}
\author[HKU]{Lijun Jiang}
\ead{jianglj@hku.hk}
\author[UESTC]{Maojun Li}
\ead{limj@cqu.edu.cn}
\author[UESTC]{Liwei Xu \corref{cor1}}
\ead{xul@uestc.edu.cn}
\cortext[cor1]{Corresponding author}

\address[CQU]{\leftline{\scriptsize College of Mathematics and Statistics, Chongqing University, Chongqing, 401331,   China} }
\address[PDU]{\leftline{\scriptsize School of Electrical and Computer Engineering, Purdue University, West Lafayette,IN 47907, USA}}
\address[HKU]{\leftline{\scriptsize Department of Electrical and Electronic Engineering, The University of Hong Kong, Hong Kong}}
\address[UESTC]{\leftline{\scriptsize School of Mathematical Sciences, University of Electronic Science and Technology of China,  Sichuan 611731,  China}}

\begin{abstract}
It is known that both linear and nonlinear optical phenomena can be produced when the plasmon in metallic nanostructures are excited by the external electromagnetic waves. In this work, a coupled system of Maxwell equations and a gas dynamic model including a quantum pressure term is employed to simulate the plasmon dynamics of free electron fluid in different metallic nanostructures using a discontinuous Galerkin method in two dimensions.  Numerical benchmarks demonstrate that the proposed numerical method can simulate both the high order harmonic generation and the nonlocal effect from  metallic nanostructures. Based on the switch-on-and-off investigation, we can conclude that the quantum pressure term in gas dynamics is responsible for the bulk plasmon resonance. In addition, for the dielectric-filled nano-cavity, a coupled effective polarization model is further adopted to investigate the optical behavior of bound electrons. Concerning the numerical setting in this work, a strengthened influence of bound electrons on the generation of high order harmonic waves has been observed.
\end{abstract}


\end{frontmatter}

\section{Introduction}
\label{sec:1}
In recent years, there are increasing interests in complex optical phenomena associated with metallic nanostructures.  One of them is the classical local optical response that features the macroscopic properties of materials.  However, for small metallic nanoparticles and metallic clusters, the experimental studies \cite{Ciraci2012} have retrieved a size-dependent surface resonance shift and a multiple bulk resonance at frequencies above the plasma frequency in the  extinction cross sections (ECS).
These novel phenomena, which are not observed in the classical local response, are due to the instantaneous response to the excitation in a nonlocal manner:  $\bold D(\bm{x},\omega)=\epsilon_0\int\bm{\epsilon}(\bm{x},\bm{x}^{'},\omega)\cdot\bold E(\bm{x}^{'},\omega)d\bm{x}^{'}$ , and thus are termed as the so-called nonlocal effect. Here,  $\bold D$ is the electric displacement, $\epsilon_0$ is the electric permittivity in vacuum, $\bm{\epsilon}$ is the relative permittivity tensor, $\omega$ is the frequency, and $\bold E$ denotes the electric field.
 These plasmonic responses have found their wide applications in biosensing \cite{Anker2008}, plasmonic waveguiding \cite{Bozhevolnyi2006} and cancer therapy \cite{Lal2008}.
Another attractive optical phenomenon is the high order harmonic generation, in particular, the second harmonic generation (SHG). Physically, the SHG is an optical process in which an electromagnetic wave at the fundamental frequency interacts with metallic nanoparticles to generate a new wave with twice of the fundamental frequency.  Experiments \cite{Klein2006,Klein2007,Niesler2009,Sonnenberg1968} have shown that the second harmonic wave can be generated from various metamaterials. The SHG is of great importance with broad applications such as in development of the laser sources \cite{Patel1965}, the optical parametric amplifiers \cite{Wasylczyk2005}, and imaging and microscopy technology \cite{Chen2012}.

There have been many numerical methods and models proposed to simulate the above optical phenomena in nanostructures.   In \cite{Liu2010}, the authors generalized a finite-difference time domain (FDTD) method to simulate the SHG from metallic nanostructures by using a fully coupled fluid-Maxwell system (called the nonlinear hydrodynamic Drude model) that was derived from the cold-plasma wave equations \cite{Zeng2009} and the Maxwell  equations. In the coupled system, the charge density $\rho$ depends on the divergence of the electric field $\mathbf{E}$, namely, $\rho=\epsilon_0 \nabla \cdot \mathbf{E}$.  Since the normal electric field is discontinuous at the dielectric-metal interface, the computation of the charge density is challenging. The authors introduced a smooth transition layer between the metal and dielectric materials so that the ion density varies from its bulk value to zero smoothly, leading to an efficient computation of $\nabla \cdot \mathbf{E}$.  In \cite{Krasavin2016}, a fully second order hydrodynamic  model has been employed to explore the mechanism of nonlocal feature on the nonlinear high order harmonic generation. In \cite{Fengyan2017}, an energy stable discontinuous Galerkin method has been designed for the Maxwell equations in Kerr-Raman-type nonlinear optical media for the simulation of third harmonic wave. Numerical investigations on the high order harmonic generation using a perturbation hydrodynamic model \cite{Zeng2009}  and a fully hydrodynamic model  \cite{Fang2016,Hille2016}  are also reported in literatures.  Concerning the nonlocal effect, a mixed finite element method (FEM) adopting the N\'{e}d\'{e}lec element has been developed in \cite{Hiremath2012} for simulating the nonlocal effect of a groove and a nanowire by using a nonlocal hydrodynamic Drude model in the two-dimensional frequency domain.  In \cite{Schmitt2016}, the authors presented a DGTD method to solve a linearized nonlocal dispersion model for studying the nonlocal dispersion effect from the interaction of light with nanometer scale metallic structures.  In  \cite{Vidal2018}, the authors applied a hybridizable discontinuous Galerkin (HDG) method to solve the Maxwell equations coupled with the nonlocal hydrodynamic  Drude  model in the frequency domain for computing the nonlocal electromagnetic effect from a two-dimensional gold nanowire and a three-dimensional periodic annular nanogap structure.

In this paper, we employ a coupled system of the gas dynamic equations including the pressure term and the Maxwell equations (termed as the modified nonlinear hydrodynamic Drude (MNHD) model in the current work) to simulate both the high-order harmonic generation  and the nonlocal effect in two-dimensional metallic nanostructures using the high order Runge-Kutta discontinuous Galerkin (RKDG) method \cite{Cockburn1998}.    The gas dynamic model is essentially the Euler equations governing the motion of the electron fluid in metallic nanostructures. Being slightly different from the fully coupled fluid-Maxwell system
in \cite{Liu2010} where the high order harmonic generation has been successfully observed in numerical results, the MNHD model introduces the Thomas-Fermi pressure which characterizes the electron fluid  equation of state.  Meanwhile,  the classical linear Drude model which can be derived from the MNHD model has been applied successfully to study the nonlocal effect from small metallic nano-particles by introducing a current diffusion term \cite{Mortensen2014,Toscano2015}. These two facts indicate the potential of the MNHD model for the simulation of both the nonlocal effect  and   the high order harmonic generation under an uniform framework, and this is indeed one of the motivations of current work. To our knowledge, there have not been literatures addressing these two important optical phenomena in an uniform numerical model and method yet. Moreover, in order to simulate the SHG from a metallic nanostructure adjacent to a nanostructure
with some kind of non-metallic material, we couple the MNHD model with an effective polarization model \cite{Scalora2010} to  numerically investigate the influence of bound electrons on the generation of  high order harmonics. A strengthened influence of bound electrons on the generation of high order harmonic waves associated to the numerical setting in this work  has been successfully observed. Finally, we point out that, due to the ability of the discontinuous Galerkin (DG) method on dealing with   discontinuous physical quantities  in  simulations, we will compute directly the charge density based on the mass conservation equation other than Gauss's law.

This paper is organized as follows.
In Section 2, we introduce the mathematical governing  equations, and then  present in details the numerical scheme for solving the coupled system in Section 3.  In Section 4, numerical tests on the nonlocal effect  and the second harmonic generation are presented to show the efficient performance of the model and  numerical method. Conclusions are finally presented in Section 5.

\section{Mathematical model}
\label{sec:2}

\subsection{Maxwell equations}
The governing equations for the propagation of electromagnetic fields   are the Maxwell  equations
\begin{eqnarray}
\label{Eq:Maxwell-1}
{\mu}_0\frac{{\partial} \tilde{\mathbf H}}{{\partial} \tilde{t}}+  \tilde{\nabla} \times \tilde{\mathbf E}  &=& \mathbf 0~,\\
\label{Eq:Maxwell-2}
{\epsilon}_0\frac{{\partial} \tilde{\mathbf E}}{{\partial} \tilde{t}}- \tilde{\nabla} \times \tilde{\mathbf H} &=& - \tilde{\mathbf J}  - \tilde{\mathbf J}_b ~.
\end{eqnarray}
Here, ${\epsilon}_0,{\mu}_0$ are the permittivity and the permeability in free space, respectively,  $\tilde{\mathbf E}= \tilde{\mathbf E}^i+ \tilde{\mathbf E}^s$ and $\tilde{\mathbf H} = \tilde{\mathbf H}^i+\tilde{\mathbf H}^s$ are the total fields, with $\tilde{\mathbf E}^i$ and $\tilde{\mathbf H}^i$ being the incident fields, and $\tilde{\mathbf E}^s$ and $\tilde{\mathbf H}^s$  being the scattered fields satisfying the Silver-M$\ddot{u}$ller radiation condition
 \begin{equation}
   \label{readiation}
   \lim_{|\tilde{\mathbf x}|\rightarrow\infty}( \tilde{\mathbf H}^s \times \tilde{\mathbf x}-| \tilde{\mathbf x}| \tilde{\mathbf E}^s)=0,~ (~\textrm{or}~ \lim_{|\tilde{\mathbf x}|\rightarrow\infty}(\tilde{\mathbf E}^s\times \tilde{\mathbf x}+|\tilde{\mathbf x}|\tilde{\mathbf H}^s)=0 ),
 \end{equation}
where $\tilde{\mathbf J}$ denotes the current density generated by the motion of electrons in a metallic nanostructure and thus  identically equals to zero outside the metallic nanostructure, and $\mathbf{\tilde{J}}_b$ is the bound current density which is trivial if the effect of bound  electrons is neglected. In this paper, we take an assumption of $z$-invariance, and therefore the following notation for the curl operator is used for the vector field  $\bold u=(u_x,u_y,u_z)$:
\begin{align*}
\nabla\times \bold u &= \left(\frac{\partial u_z}{\partial y},-\frac{\partial u_z}{\partial x},\frac{\partial u_y}{\partial x} - \frac{\partial u_x}{\partial y}\right)^{\top}~.
\end{align*}

\subsection{Gas dynamic equations}

\begin{figure}[H]
  \centering
  \includegraphics[width=10cm]{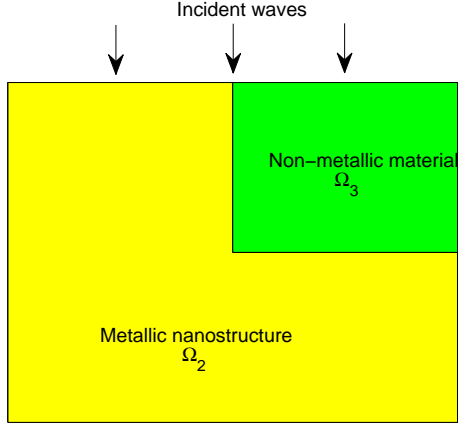}\\
  \caption{Illustration for the nanostructure}\label{Fig:nano}
\end{figure}

As being shown in Figure \ref{Fig:nano},  $\Omega_2 \subset \mathbb{R}^2$ is a bounded domain occupied by a metallic nanostructure, which is excited by an external electromagnetic field with the electric field $\tilde{\mathbf{E}}(\tilde{\mathbf x}, \tilde{t})$ and the magnetic field $\tilde{\mathbf{H}}(\tilde{\mathbf x}, \tilde{t})$.  On the other hand, the infinite mass of the ions in metal is further assumed, and it implies that the ions density $\tilde{n}_0(\bm{x})$ is a time independent function. Therefore, the positively charged ions merely play a role in providing background for the motion of  electrons without making any contribution to the current density.

As a result, in terms of continuum mechanics, the motion of  free electrons in the metallic nanostructure  $\Omega_2$   satisfies the following Euler equations
\begin{equation}\label{Eq:Euler-1}
 \frac{{\partial} (\tilde{n}_e \tilde{q}_e)}{{\partial} \tilde{t} }+ \tilde{\nabla} \cdot(\tilde{n}_e \tilde{q}_e \tilde{\mathbf u}_e) = 0~,
\end{equation}
 \begin{equation}\label{Eq:Euler-2}
   \tilde{m}_e\left[\frac{ {\partial} \tilde{\mathbf u}_e}{ {\partial} \tilde{t}}+( \tilde{\mathbf u}_e \cdot \tilde{\nabla}) \tilde{\mathbf u}_e\right] = \tilde{q}_e(\tilde{\mathbf E}+ {\mu}_0 \tilde{\mathbf u}_e \times \tilde{\mathbf H})- \tilde{\gamma} \tilde{m}_e \tilde{\mathbf u}_e-\frac{ \tilde{\nabla} \tilde{p}}{\tilde{n}_e}~,
\end{equation}
where $\tilde{q}_e, \tilde{m}_e, \tilde{n}_e, \tilde{\mathbf u}_e$ denote the electron charge, mass, number density and velocity field, respectively. $\tilde{\gamma}=1/\tilde{\tau}$ denotes the  relaxation time, the average time of collisions between the electrons and the ions. $\tilde{p}={(3\pi^2)}^{2/3}(\hbar^2/5 \tilde{m}_e) \tilde{n}_e^{5/3}$, with $\hbar$ being the Planck's constant, is the electron gas pressure evaluated by the Thomas-Fermi theory \cite{Parr1994}. In this work, we term the equations \eqref{Eq:Maxwell-1}-\eqref{Eq:Euler-2} as the modified nonlinear hydrodynamic Drude (MNHD) model.  Since the coupling MNHD system is a self-consistent model with strong nonlinearity and diffusion effects, one expects to be able to observe both the high-order harmonic generation and the nonlocal optical response in numerics using this model.

\subsection{Polarization model}
Let $\Omega_3\subset \mathbb{R}^2$ (see Figure \ref{Fig:nano}) be a bounded domain filled with some kind of non-metallic material, being adjacent to $\Omega_2$. In the domain $\Omega_3$,  the electrons are subject to the linear and nonlinear restoring forces and can not leave far away from their atomic nucleus. We call these electrons as the bound electrons.  Let $\mathbf{\tilde P}$ be the polarization and $\mathbf{\tilde{J}}_b$ be the bound current density, respectively, an effective polarization model for bound charges can be obtained by using Newton's second law and neglecting the nonlinear restoring forces \cite{Scalora2010}
\begin{eqnarray}\label{Eq:bound-1}
\frac{{\partial} \mathbf{\tilde P} }{{\partial} \tilde{t}} = \mathbf{\tilde{J}_b},\\ \label{Eq:bound-2}
\frac{{\partial} \mathbf{\tilde{J}_b} }{{\partial} \tilde{t}}+  \tilde{\gamma}_b \mathbf{\tilde{J}_b}  +\tilde{\omega}_b^2 {\mathbf{\tilde P}} &=& \frac{\tilde{n}_b \tilde{q}_e^2}{\tilde{m}_e} \tilde{\mathbf{E}}  +  \frac{{\mu}_0 \tilde{q}_e}{\tilde{m}_e {c}}  \mathbf{\tilde{J}_b} \times \mathbf{\tilde{H}}~.
\end{eqnarray}
Here, $\tilde{n}_b$, $\tilde{\gamma}_b$, $\tilde{\omega}_b$  denote the constant ion density, damping coefficient and resonance frequency for the bound electrons, respectively,  and ${c}$ denotes the speed of light in vacuum. We point out that this model defined on the domain $\Omega_3$ is coupled with the MNHD  model for the numerical simulation only when we consider the numerical test on the influence of bound electrons (the last part in Section 4.2).

\subsection{Compact nondimensionalized forms}
Assume that $\Omega_1\subset\mathbb{R}^2$ is a finite truncation  such that $(\Omega_2\cup\Omega_3) \subset \Omega_1$.  Defining the charge density $\tilde{\rho}$ and the current density $\tilde{\mathbf J}$ as follows
\begin{equation}
\nonumber
  \tilde{\rho} =  \tilde{n}_e \tilde{q}_e~,  \qquad
  \tilde{\mathbf J}  =  \tilde{\rho} \tilde{\mathbf u}_e~,
\end{equation}
 we can write the equations \eqref{Eq:Maxwell-1}-\eqref{Eq:Maxwell-2} and \eqref{Eq:Euler-1}-\eqref{Eq:Euler-2}  in the following compact hyperbolic system   after a nondimensionalization procedure according to Table \ref{Tab:units system}
\begin{equation}\label{Eq:conservation1}
  \frac{\partial U_1}{\partial t}+\nabla\cdot\bm{F}_1(U_1)=S_1(U_2,U_3),~ \textmd{in}~\Omega_1\times[0,T]~ ,
\end{equation}
\begin{equation}\label{Eq:conservation2}
  \frac{\partial U_2}{\partial t}+\nabla\cdot\bm{F}_2(U_2)=S_2(U_1,U_2),~\textmd{in}~\Omega_2\times[0,T],
\end{equation}
 where $\bm{F}_1(U_1)=\left(F_1(U_1),G_1(U_1)\right)$, $\bm{F}_2(U_2)=\left(F_2(U_2),G_2(U_2)\right)$. We could also formulate the  polarization model into the  following compact  ordinary differential system according to Table \ref{Tab:units system}
\begin{equation}\label{Eq:conservation3}
  \frac{d U_3}{d t}=S_3(U_1,U_3),~\textmd{in}~\Omega_3\times[0,T]~.
\end{equation}
In the above formulations, \[U_1 = \left( \begin{array}{l}
H_x\\
H_y\\
H_z\\
E_x\\
E_y\\
E_z
\end{array} \right),
F_1(U_1) = \left( \begin{array}{l}
~~~0~~\\
-E_z~\\
~~E_y\\
~~~0~\\
~~H_z\\
-H_y~
\end{array} \right),
G_1(U_1) = \left( \begin{array}{l}
~~E_z~~\\
~~~0~~~\\
-E_x~~~\\
-H_z\\
~~~0~~~\\
~~H_x
\end{array} \right),
S_1 = -\left( \begin{array}{l}
0\\
0\\
0\\
\rho u_x + J_{bx}\\
\rho u_y + J_{by}\\
\rho u_z + J_{bz}\\
\end{array} \right)
\]
\[U_2 = \left( \begin{array}{l}
\rho\\
\rho u_x\\
\rho u_y\\
\rho u_z
\end{array} \right),
F_2(U_2) = \left( \begin{array}{l}
\rho u_x\\
\rho u_xu_x+k\rho^{5/3}\\
\rho u_xu_y\\
\rho u_xu_z
\end{array} \right),
G_2(U_2) = \left( \begin{array}{l}
\rho u_y\\
\rho u_xu_y\\
\rho u_yu_y+k\rho^{5/3}\\
\rho u_xu_z
\end{array} \right),\]
\[k=\frac{1}{5}{\left(\frac{\hbar}{m_e}\right)}^2{\left(\frac{3\pi ^2}{q_e}\right)}^{2/3}~,
S_2 = \left( \begin{array}{l}
0\\
\frac{\rho q_e}{m_e}(E_x+ u_yH_z- u_zH_y)-\gamma\rho u_x\\
\frac{\rho q_e}{m_e}(E_y+ u_zH_x- u_xH_z)-\gamma\rho u_y\\
\frac{\rho q_e}{m_e}(E_z+ u_xH_y- u_yH_x)-\gamma\rho u_z\\
\end{array} \right)~,
\]
\[U_3 = \left( \begin{array}{l}
P_x\\
P_y\\
P_z \\
J_{bx}\\
J_{by}\\
J_{bz}
\end{array} \right)~,
S_3 = -\left( \begin{array}{l}
J_{bx}\\
J_{by}\\
J_{bz}\\
\frac{n_bq_e^2}{m_e}E_x + \frac{q_e}{m_e}(J_{by}H_z- J_{bz}H_y)-\gamma_b J_{bx}-\omega_b^2 P_x\\
\frac{n_bq_e^2}{m_e}E_y + \frac{q_e}{m_e}(J_{bz}H_x- J_{bx}H_z)-\gamma_b J_{by}-\omega_b^2 P_y\\
\frac{n_bq_e^2}{m_e}E_z + \frac{q_e}{m_e}(J_{bx}H_y- J_{by}H_x)-\gamma_b J_{bz}-\omega_b^2 P_z
\end{array} \right)~,
\]
 $T$  is  the final simulation time, and the subscript $x$ ($y$ or $z$) denotes the $x$ ($y$ or $z$) component of corresponding unknowns. Finally, we indicate that the radiation condition \eqref{readiation} will be replaced with an approximate boundary condition on the boundary $\partial\Omega_1$, which is to be discussed  in the next section.

\begin{table}[ht]\tiny\renewcommand\arraystretch{1.0}
\caption{Unit system}
\centering\label{Tab:units system}
\begin{tabular}{|c|c|c|}
  \hline
  & &\\
  Physical quantity & Reference scale & Redefined quantity \\
   \cline{1-3}
   & &\\
   $\tilde{L}$ & $L_0=1.0\times10^{-9}m$ & $L=\tilde{L}/L_0$ \\
   \cline{1-3}
   & &\\
   $\tilde{t}$ & $t_0=L_0/{c}, {c}=1/\sqrt{{\mu}_0{\epsilon}_0}$ & $t=\tilde{t}/t_0$ \\
   \cline{1-3}
   & &\\
  $\tilde{\nabla}$ &   & $\nabla=L_0\tilde{\nabla}$ \\
  \cline{1-3}
  & &\\
  ${{\partial}}/{{\partial} \tilde{t}}$ &   & ${{\partial}}/{{\partial} {t}}=t_0{{\partial}}/{{\partial} \tilde{t}}$ \\
  \cline{1-3}
  & &\\
   $\tilde{\mathbf{E}}$ & $E_0=1.0\times10^7V/m$ & $\mathbf{E}=\tilde{\mathbf{E}}/E_0$ \\
  \cline{1-3}
  & &\\
  $\tilde{\mathbf{H}}$ & $H_0=E_0/{Z}, Z=\sqrt{{\mu}_0/{\epsilon}_0}$ & $\mathbf{H}=\tilde{\mathbf{H}}/H_0$ \\
  \cline{1-3}
    & &\\
  $\tilde{\mathbf{J}}$ & $J_0=E_0/({Z}L_0)$ & $\mathbf{J}=\tilde{\mathbf{J}}/J_0$ \\
    \cline{1-3}
    & &\\
  $\tilde{\mathbf{J}}_b$ & $J_0$ & $\mathbf{J}_b=\tilde{\mathbf{J}}_b/J_0$ \\
      \cline{1-3}
      & &\\
  $\tilde{\mathbf{P}}$ & $t_0J_0$ & $\mathbf{P}=\tilde{\mathbf{P}}/(t_0J_0)$ \\
  \cline{1-3}
  & &\\
  $\tilde{\rho}$ & $\rho_0={\epsilon}_0/L_0$ & $\rho=\tilde{\rho}/\rho_0$ \\
  \cline{1-3}
  & &\\
  $\tilde{\omega}$ & $\omega_0={c}/L_0$ & $\omega=\tilde{\omega}/\omega_0$ \\
  \hline
\end{tabular}
\end{table}

\section{Numerical schemes}
\label{sec:3}
In this section, we will present the numerical method solving the equations \eqref{Eq:conservation1}-\eqref{Eq:conservation3}. Let $\mathscr{T}_h$ be a partition of $\Omega_1$. For each element $K\in\mathscr{T}_h$, we define the following finite dimensional discrete spaces consisting of piecewise polynomials with the degree at most $k$
\begin{eqnarray}
\nonumber
   V_{h,\Omega_1}^p:&=&\{U\in\left(L^2(\Omega_1)\right)^p:U|_{K}\in \left(P^k(K)\right)^p,\forall K\in\mathscr{T}_h\}~,\\
\nonumber
   W_{h,\Omega_2}^q:&=&\{U\in\left(L^2(\Omega_2)\right)^q:U|_{K}\in \left(P^k(K)\right)^q,\forall K\in \tilde{\mathscr{T}}_h\}~,\\
   \nonumber
   V_{h,\Omega_3}^p:&=&\{U\in\left(L^2(\Omega_3)\right)^p:U|_{K}\in \left(P^k(K)\right)^p,\forall K\in\hat{\mathscr{T}}_h\}~,
\end{eqnarray}
where $\tilde{\mathscr{T}}_h=\{  K \in {\mathscr{T}}_h:     K\subset \Omega_2    \}$, and $\hat{\mathscr{T}}_h=\{  K \in {\mathscr{T}}_h:     K\subset \Omega_3 \}$. We assume that the boundary of each subdomain $\Omega_i$, $ i = 1, 2, 3$, belongs to the set of boundary of $K$, or contains the vertex nodes of $K$.

\subsection{Schemes with the forward Euler time discretization}
\label{sec:3.1}
We start introducing the schemes with the first order forward Euler method for the time discretization, and the higher order time discretization  will be discussed in Section \ref{sec:3.2}. The proposed schemes evolve the numerical solutions $U_{1h}$, $U_{2h}$ and $U_{3h}$, which are assumed to be available at $t=t^n$, denoted by $U_{1h}^{n} \in V_{h,\Omega_1}^6$, $U_{2h}^{n} \in W_{h,\Omega_2}^4$ and $U_{3h}^{n} \in V_{h,\Omega_3}^6$, and will be computed at $t=t^{n+1}=t^n+\Delta t^n$, denoted by $U_{1h}^{n+1}$, $U_{2h}^{n+1}$ and $U_{3h}^{n+1}$.

\subsubsection{Updating $U_{1h}^{n+1}$}
To get $U_{1h}^{n+1}$, we apply to \eqref{Eq:conservation1} with the DG method for the space discretization and the first order forward Euler method for the time discretization. That is, to look for $U_{1h}^{n+1}\in V_{h,\Omega_1}^6$, for $\forall \,\Phi_h \in V_{h,\Omega_1}^6$ and  $\forall K \,\in \mathscr{T}_h$, such that
\begin{eqnarray}\label{SMDG1}
  \int_K U_{1h}^{n+1} \cdot \Phi_h d\bm{x} &=& \int_K U_{1h}^{n} \cdot \Phi_h d\bm{x} + \Delta t^n \int_{K}\mathbf F_1(U_{1h}^{n}) \cdot \nabla \Phi_hd\bm{x} \nonumber \\
  &-& \Delta t^n \int_{\partial K} \mathcal{H}_1(U_{1h}^{n,int},U_{1h}^{n,ext}) \cdot \Phi_hdS+ \Delta t^n \int_K S_1(U_{2h}^{n},U_{3h}^{n}) \cdot \Phi_h {d\bm x}~,
\end{eqnarray}
where ${\mathcal{H}_1(\cdot,\cdot)}$ denotes the numerical flux  evaluated on the interface between two adjacent elements, and $U_{1h}^{n,int},~ U_{1h}^{n,ext}$ are the traces of $U_{1h}^{n}$ on $\partial K$ evaluated from the interior and exterior of element $K$. In this paper, we employ the upwind numerical flux \cite{UPML} given by
\begin{equation}
\nonumber
\mathcal{H}_1(U_{1h}^{n,int},U_{1h}^{n,ext}) = \left( \begin{array}{l}
-\bm{n}_K\times\frac{(Z\bm{H}+\bm{n}_K\times\bm E)_{h}^{n,int}+(Z\bm H-\bm n_K\times\mathbf E)_h^{n,ext}}{Z_h^{n,int}+Z_h^{n,ext}}\\ \\
\bm{n}_K\times\frac{(Y\bm{E}-\bm{n}_K\times\bm H)_h^{n,int}+(Y\bm E+\bm n_K\times\mathbf H)^{n,ext}_h}{Y_h^{n,int}+Y_h^{n,ext}}
\end{array} \right)~.
\end{equation}
where $Z=\frac{1}{Y}=\sqrt{\frac{\mu}{\epsilon}}$ denotes the local impedance, and $\bm n_K$ denotes the unit outward normal of $K$.

\subsubsection{Updating $U_{2h}^{n+1}$}
To get $U_{2h}^{n+1}$, we apply to \eqref{Eq:conservation2} with the DG method for the space discretization and the first order forward Euler method for the time discretization. That is, to look for $U_{2h}^{n+1}\in W_{h,\Omega_2}^4$, for $\forall\, \Psi_h \in W_{h,\Omega_2}^4$ and $\forall K \,\in \tilde{\mathscr{T}}_h$, such that
\begin{eqnarray}\label{SMDG2}
  \int_K U_{2h}^{n+1} \cdot \Psi_h d\bm{x} &=& \int_K U_{2h}^{n} \cdot \Psi_h d\bm{x} + \Delta t^n \int_{K}\mathbf F_2(U_{2h}^{n}) \cdot \nabla \Psi_hd\bm{x} \nonumber \\
  &-& \Delta t^n \int_{\partial K} \mathcal{H}_2(U_{2h}^{n,int},U_{2h}^{n,ext}) \cdot \Psi_hdS+ \Delta t^n \int_K S_2(U_{1h}^{n},U_{2h}^{n}) \cdot \Psi_h {d\bm x}~,
\end{eqnarray}
where $\mathcal{H}_2(\cdot,\cdot)$ denotes the numerical flux evaluated on the interface between two adjacent elements, and $U_{2h}^{n,int},~ U_{2h}^{n,ext}$ are the traces of $U_{2h}^n$ on $\partial K$ evaluated from the interior and exterior of element $K$. In this paper, we employ the Lax-Friedrichs numerical flux given by

\begin{equation}
\nonumber
  \mathcal{H}_2(U_{2h}^{n,int},U_{2h}^{n,ext})=\frac{1}{2}\left[\bm F_2(U_{2h}^{n,int})\cdot\bm n_K+\bm F_2(U_{2h}^{n,ext})\cdot\bm n_K-\alpha^n\left(U_{2h}^{n,ext}-U_{2h}^{n,int}\right)\right],\alpha^n=\max_{K}|\bm F_2'(U_{2h}^{n})\cdot\bm n_K|~.
\end{equation}

\subsubsection{Updating $U_{3h}^{n+1}$}
We look for $U_{3h}^{n+1}\in V_{h,\Omega_3}^6$, for $\forall \,\xi_h \in V_{h,\Omega_3}^6$ and $\forall\, K \in \hat{\mathscr{T}}_h$, such that
\begin{eqnarray}\label{SMDG1}
  \int_K U_{3h}^{n+1} \cdot \xi_h d\bm{x} &=& \int_K U_{3h}^{n} \cdot \xi_h d\bm{x}  + \Delta t^n \int_K S_3(U_{1h}^{n},U_{3h}^{n}) \cdot \xi_h {d\bm x}~.
\end{eqnarray}

\subsection{Schemes  with high order time discretizations }
\label{sec:3.2}
In the previous subsection, we  have discussed the first order time discretization. To increase the accuracy in the time domain,  strong stability preserving (SSP) high-order time discretizations \cite{Gottlieb2001} can be used, and  we employ the third order TVD Runge-Kutta method \cite{Cockburn1998}  for the time discretization in this work.

\subsection{Numerical ingredients}
\label{sec:3.3}
Numerical investigations on complex optical phenomena are significantly correlated with numerical settings.   In this subsection,  we will describe in details numerical ingredients, including the initial conditions, and the interface and absorbing  boundary conditions, etc..

\subsubsection{Initial conditions}
\label{sec:3.3.1}
Before excited by the electromagnetic fields, the electrons in the nanostructures are at rest if the thermal effect is ignored. Under this circumstance, the electron number density is equal to the ion number density $n_0(\bm{x})$ so that   nanostructures are electrically neutral, where $n_0(\bm{x})$ could be evaluated via the plasma frequency  $\omega_p=\sqrt{\frac{n_0q_e^2}{m_e}}$.  Therefore, the initial conditions are set as follows
\begin{align}\label{initial conditions}
\nonumber
  &\mathbf E(\bm{x},0)=\mathbf H(\bm{x},0)=\mathbf 0,~\textmd{in}~\Omega_1~,\\\nonumber
  &\rho(\bm{x},0)=q_en_0(\bm{x}),~ \mathbf u(\bm{x},0)=\mathbf 0,~\textmd{in}~\Omega_2~,\\\nonumber
  &\mathbf P(\bm{x},0)={\mathbf J}_b(\bm{x},0)=\mathbf 0,~\textmd{in}~\Omega_3~.
\end{align}

\subsubsection{Boundary conditions on $\Gamma_2=\partial \Omega_2$}
\label{sec:3.3.2}

In order to solve the MNHD model, two boundary conditions need to be prescribed, one of which is the boundary condition on the metal-vacuum interface $\Gamma_2=\partial \Omega_2$. At the microscopic level, the charge density $\rho$ varies continuously across the dielectric-metal interface, and there is actually a transition region with a scale of a few atomic diameters where the charge density changes gradually down to be trivial \cite{Toscano2015,Liu2010}. However, the thickness of the transition layer is a negligible scale compared to the finest mesh that we are able to afford in domain discretizations. This fact makes it impossible for us to implement the ab-$initio$ boundary condition in a macroscopic model \cite{Jackson1999}. We use a natural boundary condition  in this work, namely $\frac{\partial \rho}{\partial\bm{n}}=0$. For the current density $\mathbf J$, we employ the so-called $slip$ boundary condition, i.e. $\mathbf n\cdot\mathbf J=0$. It implies that the current density is prohibited to travel out of the nanoparticle surface in a normal direction with respect to the interface while a tangential current shift is allowed.

\subsubsection{Artificial boundary conditions on $\Gamma_1=\partial \Omega_1$}
\label{sec:3.3.3}
\begin{figure}[H]
  \centering
  \includegraphics[width=8cm]{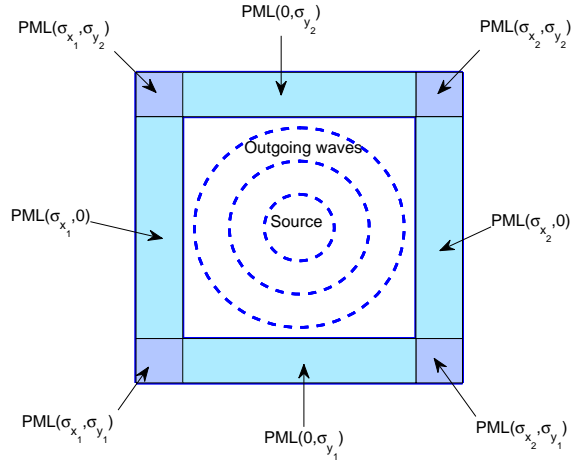}\\
  \caption{perfectly matched layer}\label{Fig:pml}
\end{figure}

As it is mentioned above,  we need to employ an artificial boundary surrounding
$\Omega_2\cup\Omega_3$, denoted by $\Gamma_1=\partial \Omega_1$,
for the practical computation.  In this paper, we use the uniaxial perfectly matched layer (PML) \cite{UPML} to absorb the electromagnetic waves propagating through the boundary $\Gamma_1$ except for those tests with particular specifications. Let $\Omega_p$ be the
PML region (see Figure \ref{Fig:pml}) surrounding the finite truncation $\Omega_1$, and
the modified formulations for \eqref{Eq:conservation1} in $\Omega_p$ are written as
\begin{equation}\label{Eq:conservationpml}
 \frac{\partial U_1}{\partial t}+\nabla\cdot\bm{F}_1(U_1)=S_p(U_1,U_7),~ \textmd{in}~\Omega_p\times[0,T]~,
\end{equation}
and
\begin{equation}\label{Eq:conservationADEpml}
  \frac{d U_7}{d t}=S_7,~\textmd{in}~\Omega_p\times[0,T]~,
\end{equation}
where
\[
S_p = \left( \begin{array}{l}
Q_x+(\sigma_x-\sigma_y)H_x\\
Q_y+(\sigma_y-\sigma_x)H_y\\
Q_z-(\sigma_x+\sigma_y)H_z\\
P_x+(\sigma_x-\sigma_y)E_x\\
P_y+(\sigma_y-\sigma_x)E_y\\
P_z-(\sigma_x+\sigma_y)E_z\\
\end{array} \right),~
U_7=
\left(\begin{array}{l}
Q_x\\
Q_y\\
Q_z\\
P_x\\
P_y\\
P_z\\
\end{array} \right),~\\
S_7 = \left( \begin{array}{l}
-\sigma_xQ_x-\sigma_x(\sigma_x-\sigma_y)H_x\\
-\sigma_yQ_y-\sigma_y(\sigma_y-\sigma_x)H_y\\
-\sigma_x\sigma_yH_z\\
-\sigma_xP_x-\sigma_x(\sigma_x-\sigma_y)E_x\\
-\sigma_yP_y-\sigma_y(\sigma_y-\sigma_x)E_y\\
-\sigma_x\sigma_yE_z\\
\end{array} \right).
\]
Here, the parameters of the dissipative layer for absorbing the fields propagating in the $i$-th direction $\sigma_i$ are given by
\begin{equation}
  \nonumber
  \sigma_i=\sigma_m\left(\frac{d_i}{\delta}\right)^n,~ i=x,y~,
\end{equation}
where $d_i, \delta, n$ denote the distance from the PML-vacuum interface, the thickness of PML, and the degree of polynomials, respectively. $\sigma_m$ is the maximum electric conductivity which can be determined by
\begin{equation}
\nonumber
  R(0)=e^{-2\sigma_m\delta/(n+1)}~,
\end{equation}
where $R(0)$ denotes the theoretical reflection at normal incidence.
The equations \eqref{Eq:conservationpml}-\eqref{Eq:conservationADEpml} are
solved by the RKDG method as well.

\subsubsection{TF/SF technique}
\label{sec:3.3.4}
\begin{figure}
  \centering
  \includegraphics[width=7.0cm]{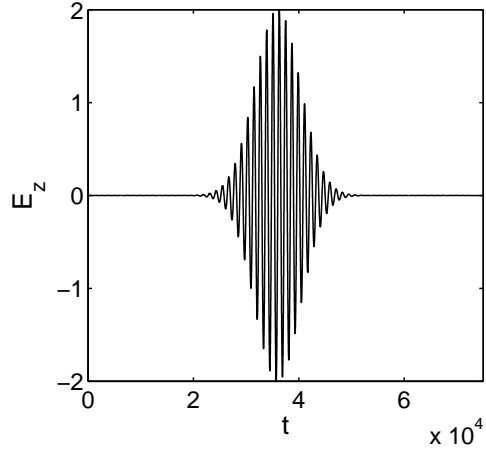}
  \caption{Gaussian pulse: ${E}_A=2.0$, ${\lambda}_0=1200$, ${t}_d=3.598\times10^4$, ${t}_b=0.6 {t}_d$.}
  \label{Fig:incidentwave}
\end{figure}

\begin{figure}
  \centering
  \includegraphics[width=7.5cm]{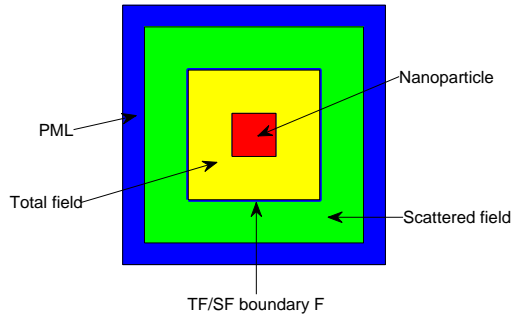}\\
  \caption{Total-field and Scattered-field}
  \label{Fig:TFSF}
\end{figure}

\begin{figure}
\centering
\includegraphics[width=7.5cm]{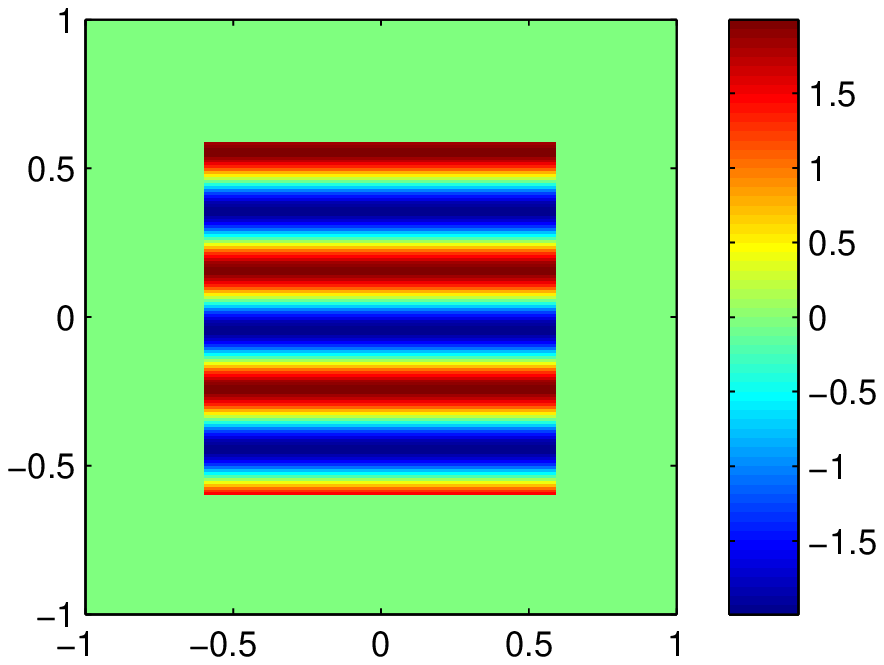}
\includegraphics[width=7.5cm]{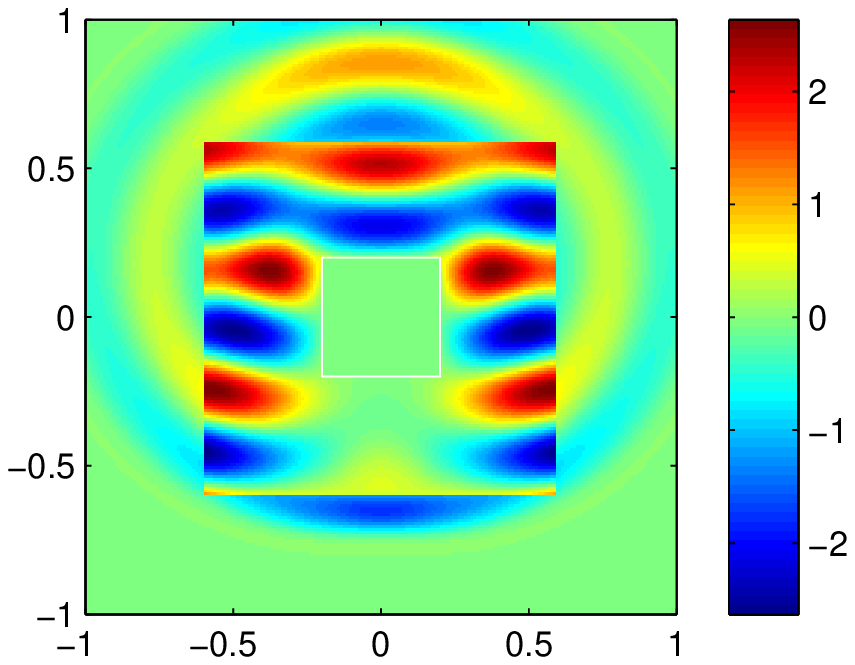}
\caption{{Injection of sine plane wave via TF/SF, Left: Electric field ${E}_z$ without object. Right: Electric field ${E}_z$ with a square perfect electric  conductor in total zone}.}
\label{Fig:tfsfDG}
\end{figure}

In order to stimulate the plasmon resonances in the metallic nanostructures, the initial conditions presented in \ref{sec:3.3.1} are not sufficient, and an extra appropriate wave source should be added during the computation. In this paper, we take the incident wave as a $z$-polarized Gaussian pulse modulated by the sine function
\begin{equation}
 \nonumber
   {E}_z={E}_A sin({\omega}_m t) e^{(-\frac{4\pi( {t}- {t}_d)^2}{ {t}_b^2})},~ {\omega}_m=\frac{{2\pi}}{{\lambda}_0},
\end{equation}
where ${\lambda}_0$ is the carrier center wavelength, ${E}_A$ is the peak amplitude, ${t}_d$ is the pulse duration, and $ {t}_b$ is the optical bandwidth (see Figure \ref{Fig:incidentwave}).
To implement this source injection in  simulations, since the popular method of hard source in computational electromagnetics may produce backward-scattered waves in a longtime simulation, we apply in this work the total-field/scattered-field (TF/SF) technique \cite{Taflove1995} which requires to divide the computational domain into a total-field zone and a scattered-field zone (see Figure \ref{Fig:TFSF}) through a virtual TF/SF boundary $F$ inside $\Gamma_1$. An incident wave is then introduced into the total-field zone from the virtual boundary without introducing any nonphysical effects. Meanwhile, a simple process on numerical fluxes along the virtual boundary  allows us to realize this purpose efficiently and accurately (see Figure \ref{Fig:tfsfDG}).

\section{Numerical results}
\label{sec:4}
 We present in this section numerical results which are computed based on the coupling system of \eqref{Eq:conservation1}-\eqref{Eq:conservation2}
  ($\bm J_b=0$ and $\Omega_3$ is empty)  except for the last case in Section \ref{sec:4.2} where we compute the MNHD model \eqref{Eq:conservation1}-\eqref{Eq:conservation2} together with the polarization model \eqref{Eq:conservation3}. In addition, all the simulations are performed with both the piecewise linear ($P^1$) and  quadratic ($P^2$) elements on structured grids, and if not specified, only $P^2$ results are presented for the illustration. The time step $\Delta t$ is dynamically determined by
\begin{align*}
\nonumber
  \Delta t =\frac{C_{cfl}}{\frac{a_x}{\Delta x}+\frac{a_y}{\Delta y}},
\end{align*}
where $a_x=\textmd{max}(|u_x|+c^f,1.0)$, $a_y=\textmd{max}(|u_y|+c^f,1.0)$, $c^f=\sqrt{\frac{5}{3}k|\rho|^{\frac{1}{3}}}$,  $C_{cfl}$ is the CFL number. If not specified, the vaule of  CFL number is selected to be $0.3$ for all tests, and we also point out that no limiting procedures have been used in all numerical tests reported in this work.

\subsection{Accuracy test}
We start with a manufactured example to study the accuracy of our schemes for MNHD model. For simplicity, we choose $k=0, \frac{q_e}{m_e}=1, \gamma=1$ in the equations \eqref{Eq:conservation1} and  \eqref{Eq:conservation2} and employ the following as the manufactured solutions
\begin{eqnarray*}\label{Eq:smooth-sol}
&\left\{
\begin{array}{lcl}
  \rho & = &1+0.5\textmd{sin}[2\pi(x+y-2t)] \\
  u_x & = & 1 \\
  u_y & = & 1 \\
  u_z & = & 0 \\
  H_x & = &  0 \\
  H_y & = & 0 \\
  H_z & = & \textmd{cos}(2\pi x)\textmd{cos}(2\pi y)\textmd{sin}(4\pi\alpha t) \\
  E_x & = & \alpha \textmd{cos}(2\pi x)\textmd{sin}(2\pi y)\textmd{cos}(4\pi\alpha t) \\
  E_y & = & -\alpha \textmd{sin}(2\pi x)\textmd{cos}(2\pi y)\textmd{cos}(4\pi\alpha t) \\
  E_z & = & 0
\end{array}
\right.
\end{eqnarray*}
where $\alpha=\frac{\sqrt2}{2}$.  The computational domain is $\Omega_2=\Omega_1=[0,1]\times[0,1]$. Periodic boundary conditions are applied in both $x-$ and $y-$directions. In Table \ref{Tab:accuracytest}, we present the $L^2$ errors and the corresponding order of accuracy for $\rho$, $H_z$, $E_x$ and $E_y$ at $T=0.2$. It shows that the orders of accuracy are optimal for both $P^k$ solutions with $k=1,2$.

\begin{table}[h]\renewcommand\arraystretch{1.0}
 \small
\caption{$L^2$ error and convergence order for $\rho$, $H_z$, $E_x$ and $E_y$ approximated by $P^1$ and $P^2$ element at $T=0.2$. }
\centering
\begin{tabular}{lllllllllllll}
  \hline\label{Tab:accuracytest}
  Mesh&&$\rho$&&&$H_z$&&&$E_x$&&&$E_y$&\\
  \cline{3-4}\cline{6-7}\cline{9-10}\cline{12-13}
   &&$L^2$ error & Order && $L^2$ error & Order && $L^2$ error & Order&& $L^2$ error & Order\\
   \cline{1-13}
   $P^1$&&&&&&&&&&&\\
   $40\times40$    &&0.106E-02&-    &&0.121E-02&-    &&0.754E-03&-    &&0.713E-03&-   \\
   $80\times80$    &&0.262E-03&2.02 &&0.299E-03&2.02 &&0.202E-03&1.90 &&0.190E-03&1.91\\
   $160\times160$  &&0.654E-04&2.00 &&0.747E-04&2.00 &&0.525E-04&1.94 &&0.491E-04&1.95\\
   $320\times320$  &&0.163E-04&2.00 &&0.187E-04&2.00 &&0.134E-04&2.00 &&0.125E-04&1.98\\
   $640\times640$  &&0.408E-05&2.00 &&0.467E-05&2.00 &&0.337E-05&1.99 &&0.315E-05&1.99\\
   $P^2$&&&&&&&&&&&\\
   $40\times40$    &&0.299E-04&-    &&0.302E-04&-    &&0.300E-04&-    &&0.297E-04&-   \\
   $80\times80$    &&0.374E-05&3.00 &&0.379E-05&2.99 &&0.383E-05&2.97 &&0.378E-05&2.97\\
   $160\times160$  &&0.467E-06&3.00 &&0.474E-06&3.00 &&0.484E-06&2.99 &&0.477E-06&2.99\\
   $320\times320$  &&0.584E-07&3.00 &&0.593E-07&3.00 &&0.608E-07&2.99 &&0.599E-07&2.99\\
   $640\times640$  &&0.730E-08&3.00 &&0.742E-08&3.00 &&0.762E-08&3.00 &&0.751E-08&3.00\\
  \hline
\end{tabular}
\end{table}

\subsection{High-order harmonic generation}
\label{sec:4.2}
\begin{figure}[H]
\centering
  {\includegraphics[height=7cm,width=5cm]{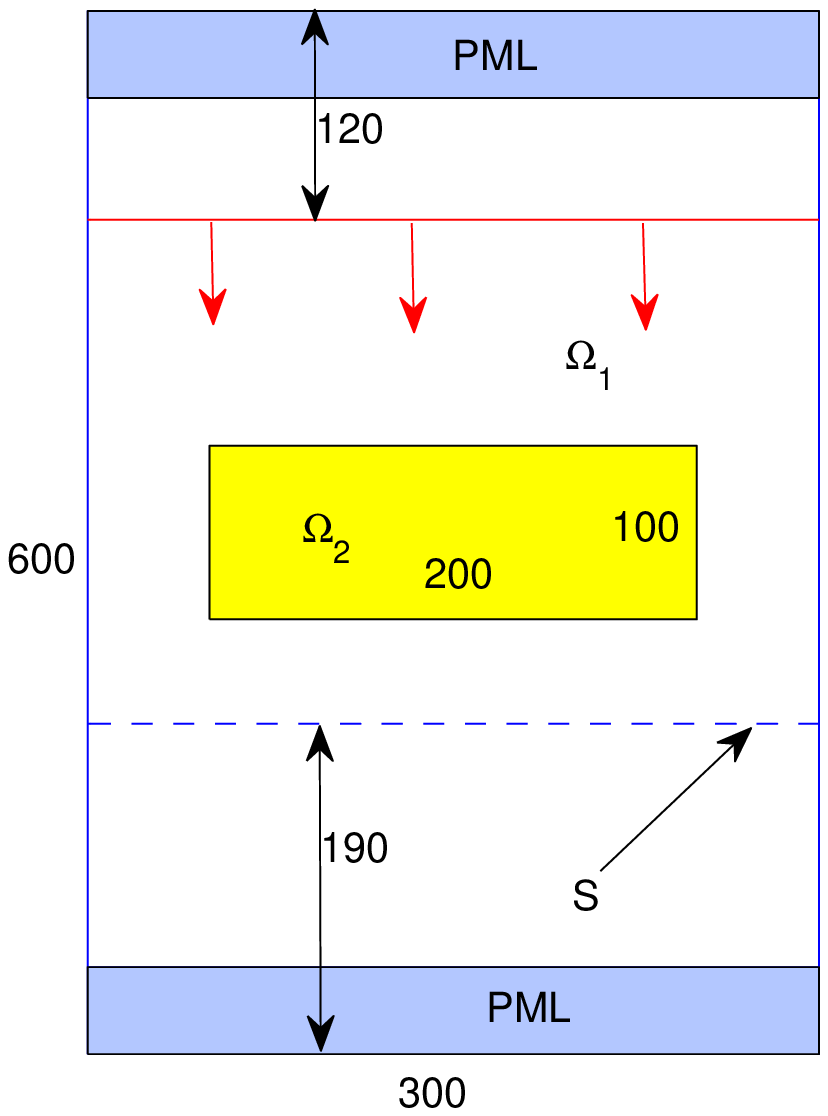}}
  {\includegraphics[height=7cm,width=5cm]{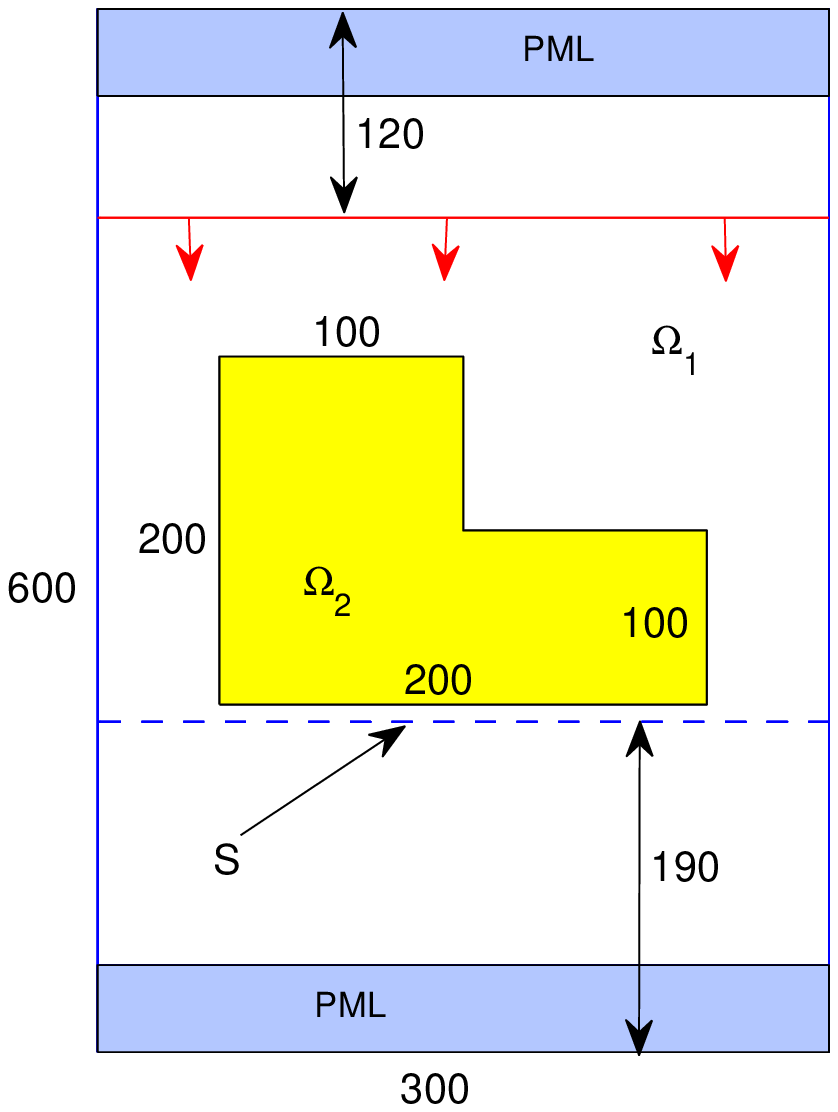}}
  {\includegraphics[height=7cm,width=5cm]{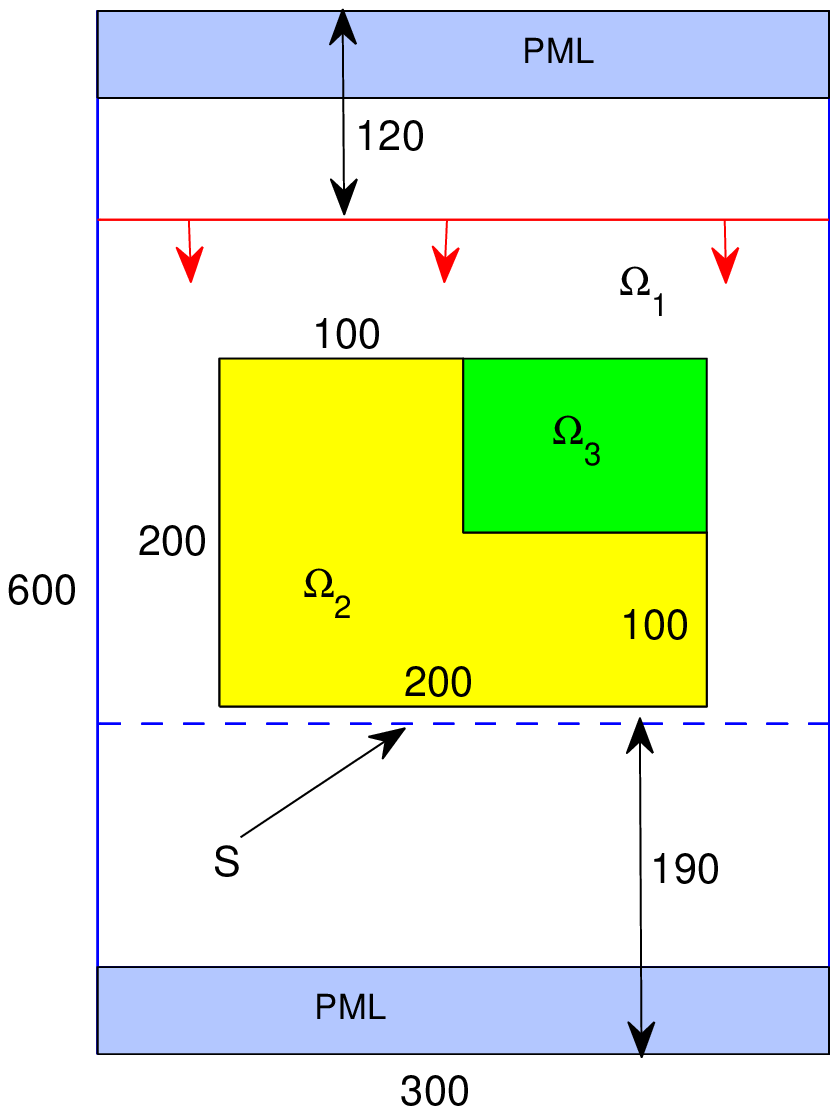}}
\caption{Setup for simulations of high-order harmonic generations.  Left: rectangular metallic nanostructure; middle: L-shaped metallic nanostructure; right: L-shaped metallic nanostructure adjacent to a rectangular non-metallic nanostructure.}
\label{setup}
\end{figure}

The high-order harmonic generation is a nonlinear optical process sensitive to the configurations of metallic nanostructures. In this test, we simulate the high-order harmonic generation from two typical metallic nanostructures, namely, an array of rectangular nanostructures (see the left of Figure \ref{setup}) and an array of L-shaped nanostructures (see the middle of Figure \ref{setup}). In addition, we also consider
an array of  L-shaped nanostructures with  metallic materials
and an array of  rectangular nanostructures with non-metallic materials (see the right of Figure \ref{setup}) for purpose of studying the influence of bound electrons on the generation of high order harmonic waves.
In our simulations, the nanostructures are arranged periodically in $x-$direction with plasma frequency ${\omega}_p=4.560\times10^{-2}$, and ${\gamma}=2.160\times10^{-4}$. For simplicity, we only focus on one single structure by using periodic boundary condition in $x-$direction and PML in $y-$direction \cite{Liu2010}.
Measurements on the linear response and the SHG signal can be computed in terms of the
integrals of the electric fields $E_z$ and $E_x$ along the probing line $S$ as follows
 \begin{eqnarray*} \label{measurement}
   \hat E_{Linear} &=& \frac{1}{|S|}\int_S E_z ds~,\\
   \hat E_{SHG} &=& \frac{1}{|S|}\int_S E_x ds~.
 \end{eqnarray*}
We employ an uniform mesh $\Delta x=\Delta y=2.0$ in the simulation and the final time $T=1.499\times10^{5}$.

\begin{figure}[H]
\centering
  {\includegraphics[width=7cm]{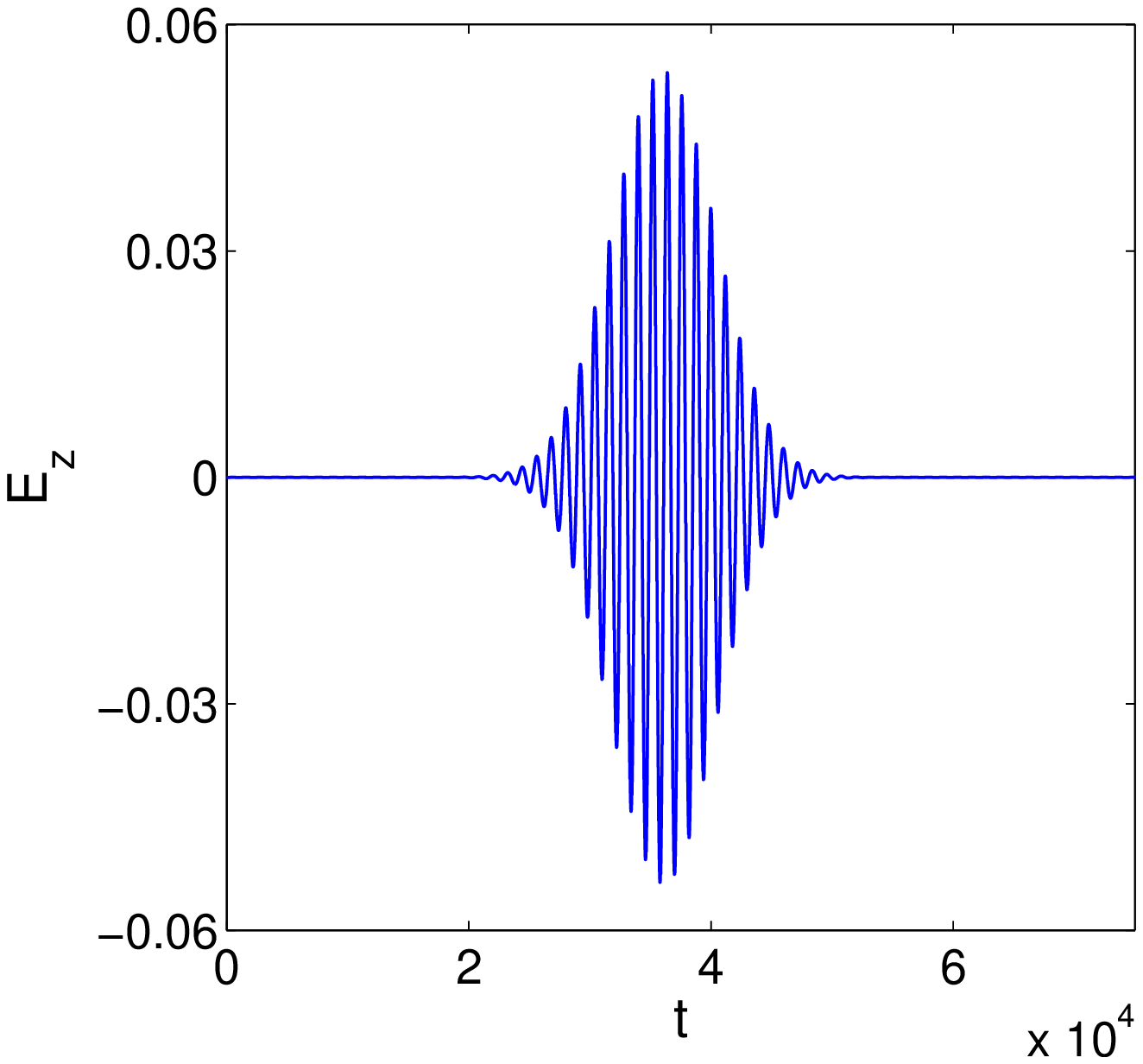}}
  {\includegraphics[width=7cm]{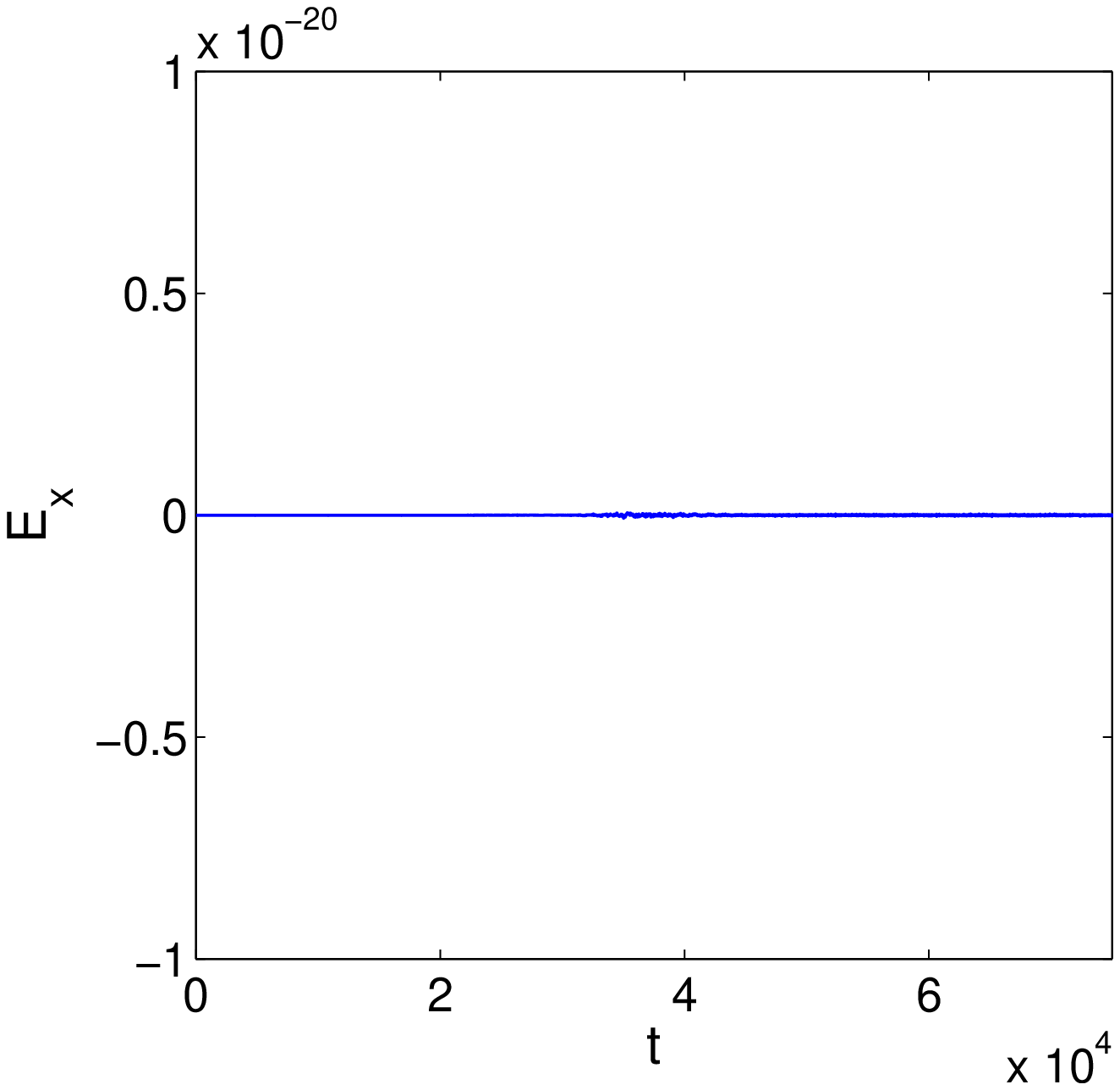}}\\
\caption{Time history of electric fields from  a rectangular nanostructure. }
\label{Rectangletime}
\end{figure}

\begin{figure}[H]
\centering
  {\includegraphics[width=7cm]{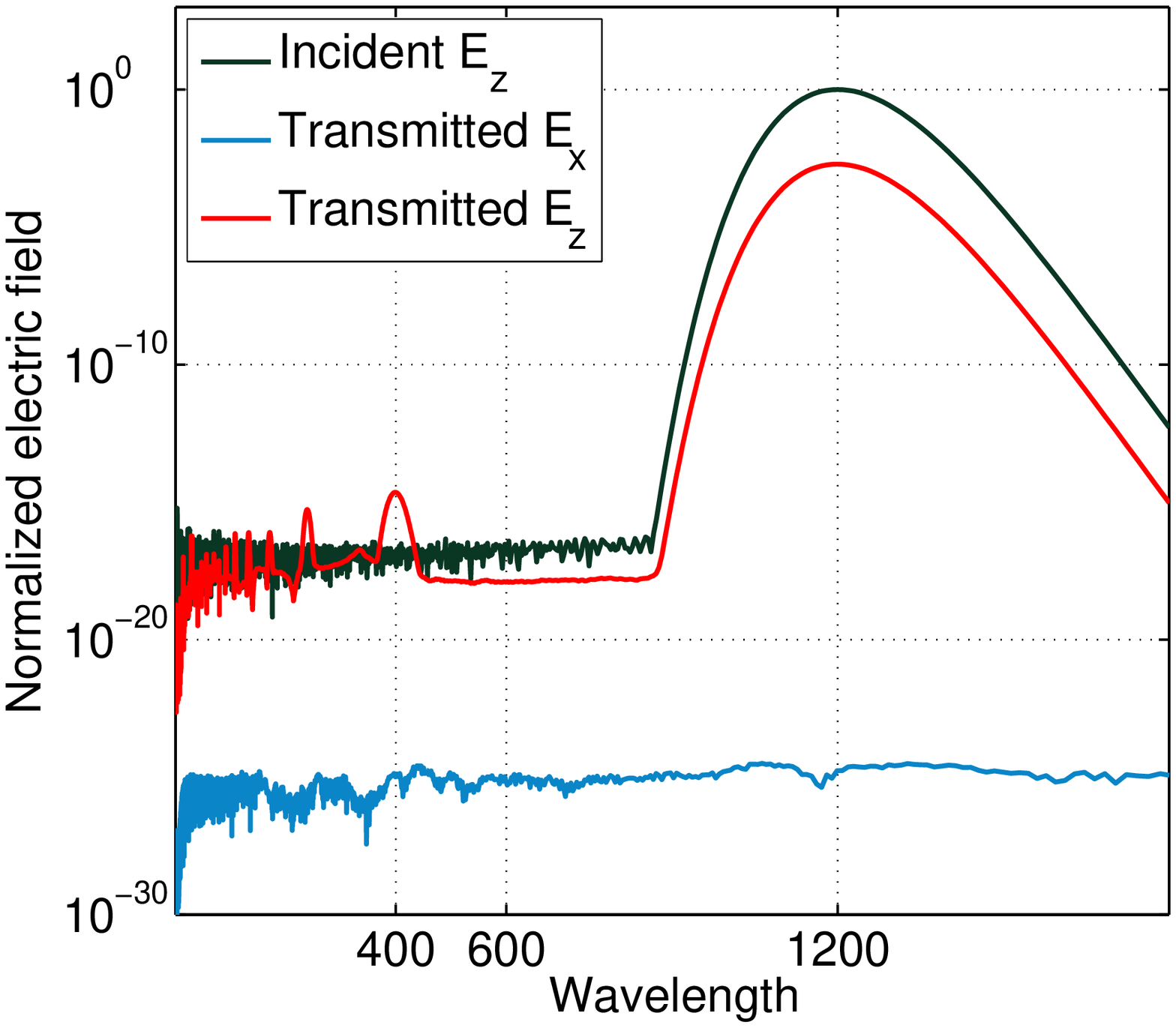}}
  {\includegraphics[width=7cm]{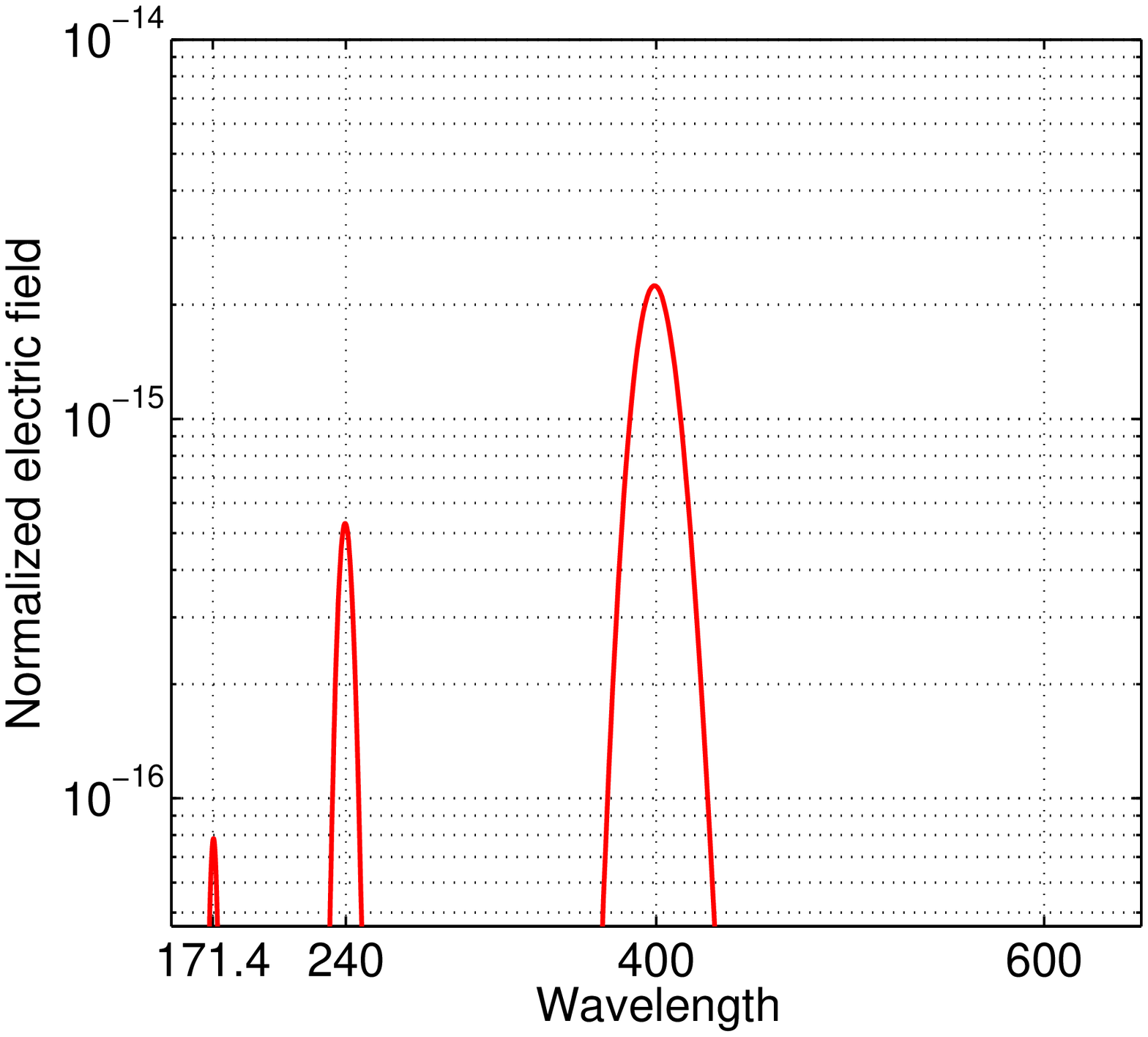}}
\caption{Spectrum with a rectangular nanostructure. Left: Spectrum of incident wave $E_{z}$, transmitted wave $E_{x}, E_{z}$; right: Zoom-in plot of the spectrum. }
\label{Rectanglefrequency}
\end{figure}

Firstly, we study the high order harmonic generation from the rectangular metallic nanostructures whose dimension is shown in Figure \ref{setup} (Left). In Figure \ref{Rectangletime}, we plot the time history of linear response and SHG. As being shown in Figure \ref{Rectangletime}, there is no SHG from the rectangular metallic particle. This result is  reasonable since the rectangular metallic nanostructure possesses a perfect centrosymmetric property which leads to a vanishing second-order nonlinear optical susceptibility tensor $\chi^{(2)}$
prohibiting the SHG. However, the rectangular metallic nanostructure does allow for the third harmonic generation (THG) since the symmetric property would not remove the third-order nonlinear optical susceptibility tensor $\chi^{(3)}$.  In Figure \ref{Rectanglefrequency} (Left), spectrums of these responses are presented. In order to show the spectrums apparently, a zoom-in plot of the spectrum is presented in Figure \ref{Rectanglefrequency} (Right). It can be observed that, the third-order harmonic generation ($400$), the fifth-order harmonic generation ($240$) and the seventh-order harmonic generation ($171.4$) are captured in the simulation. These figures also indicate that, apart from the nonlinear optical responses, the MNHD model maintains the description of the linear optical interaction.

\begin{figure}[H]
\centering
  {\includegraphics[width=7cm]{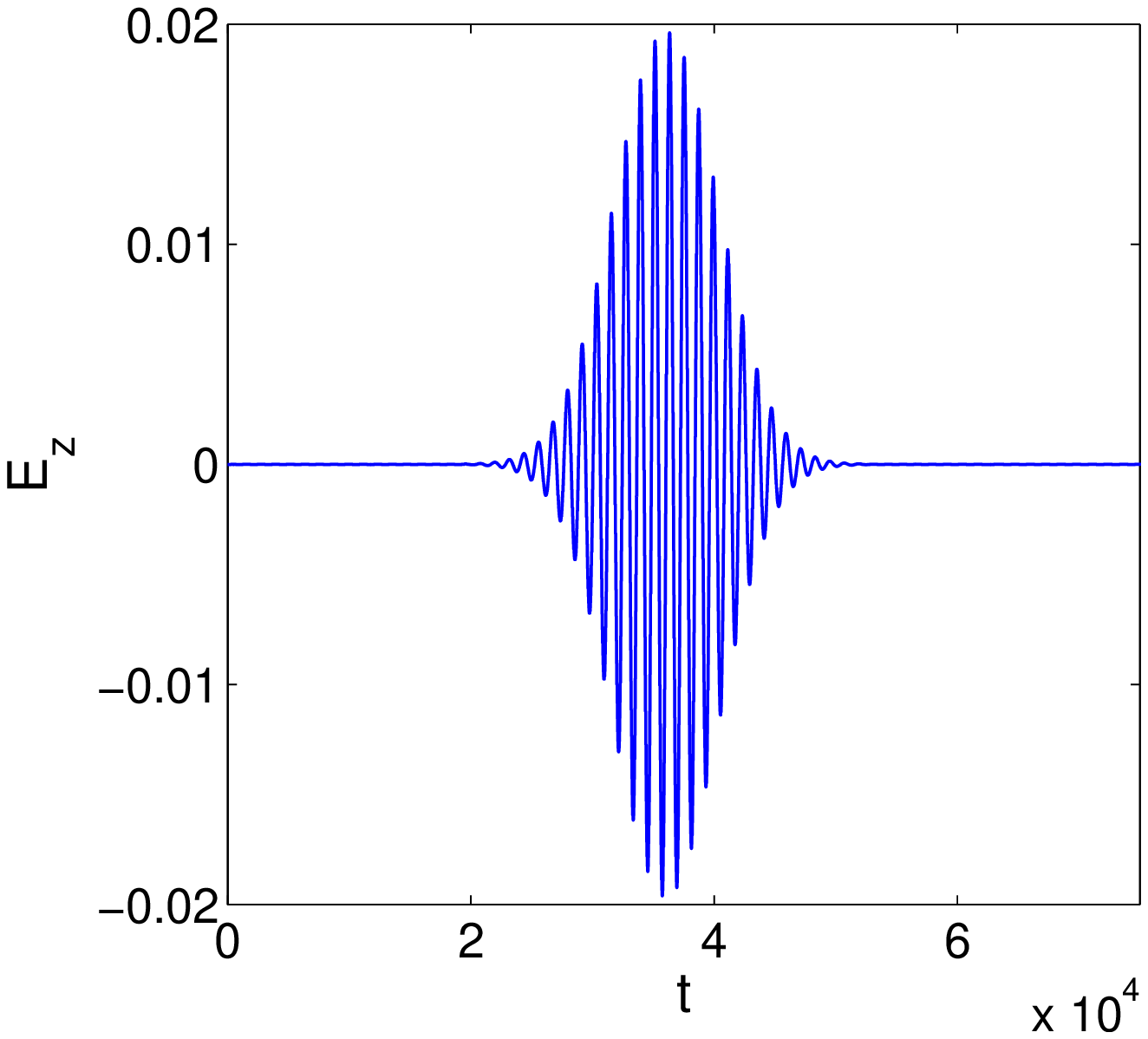}}
  {\includegraphics[width=7cm]{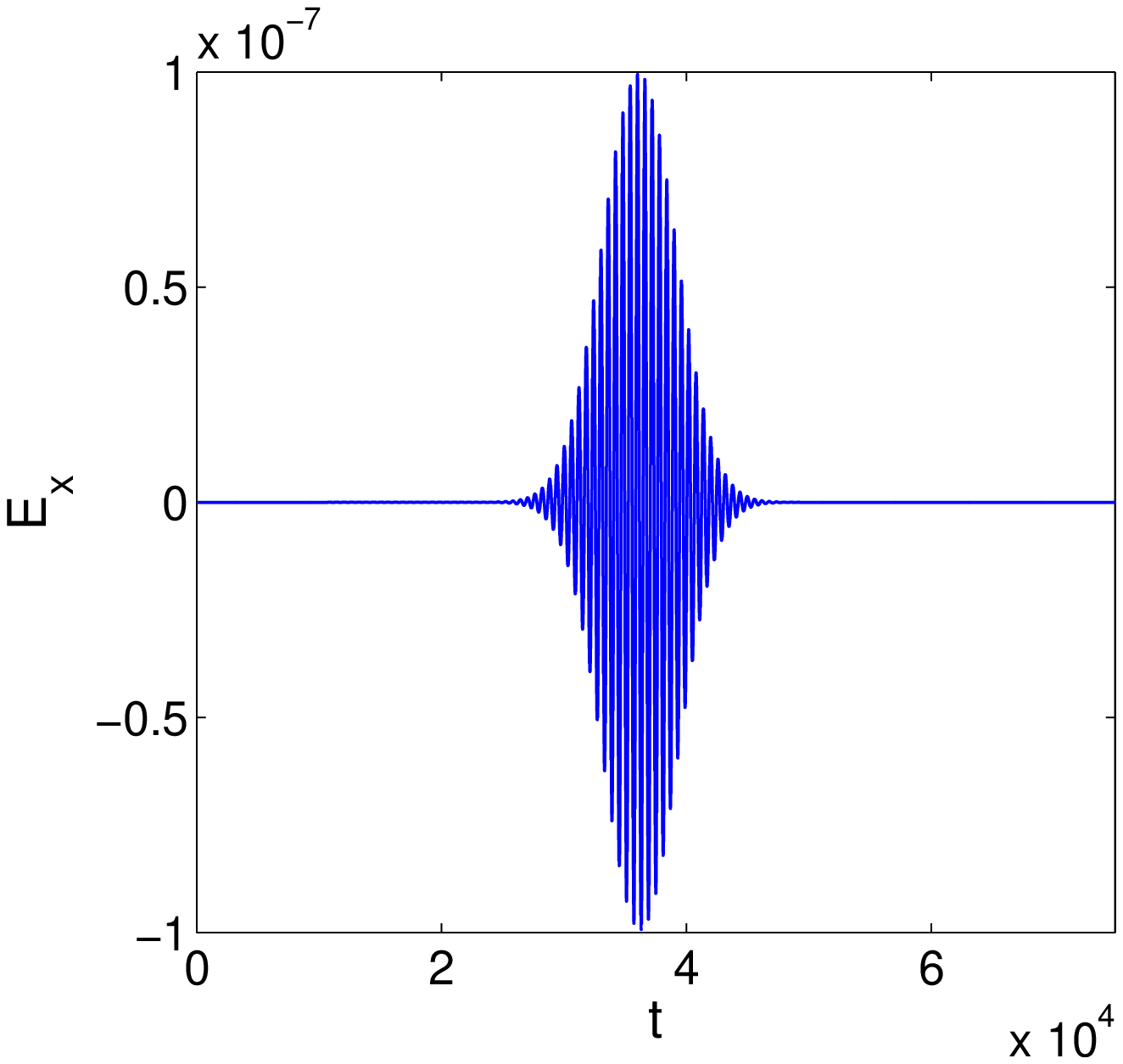}}\\
\caption{Time history of electric fields from  a L-shaped nanostructure.}\label{L-shapedtime}
\end{figure}

\begin{figure}[H]
\centering
  {\includegraphics[width=7cm]{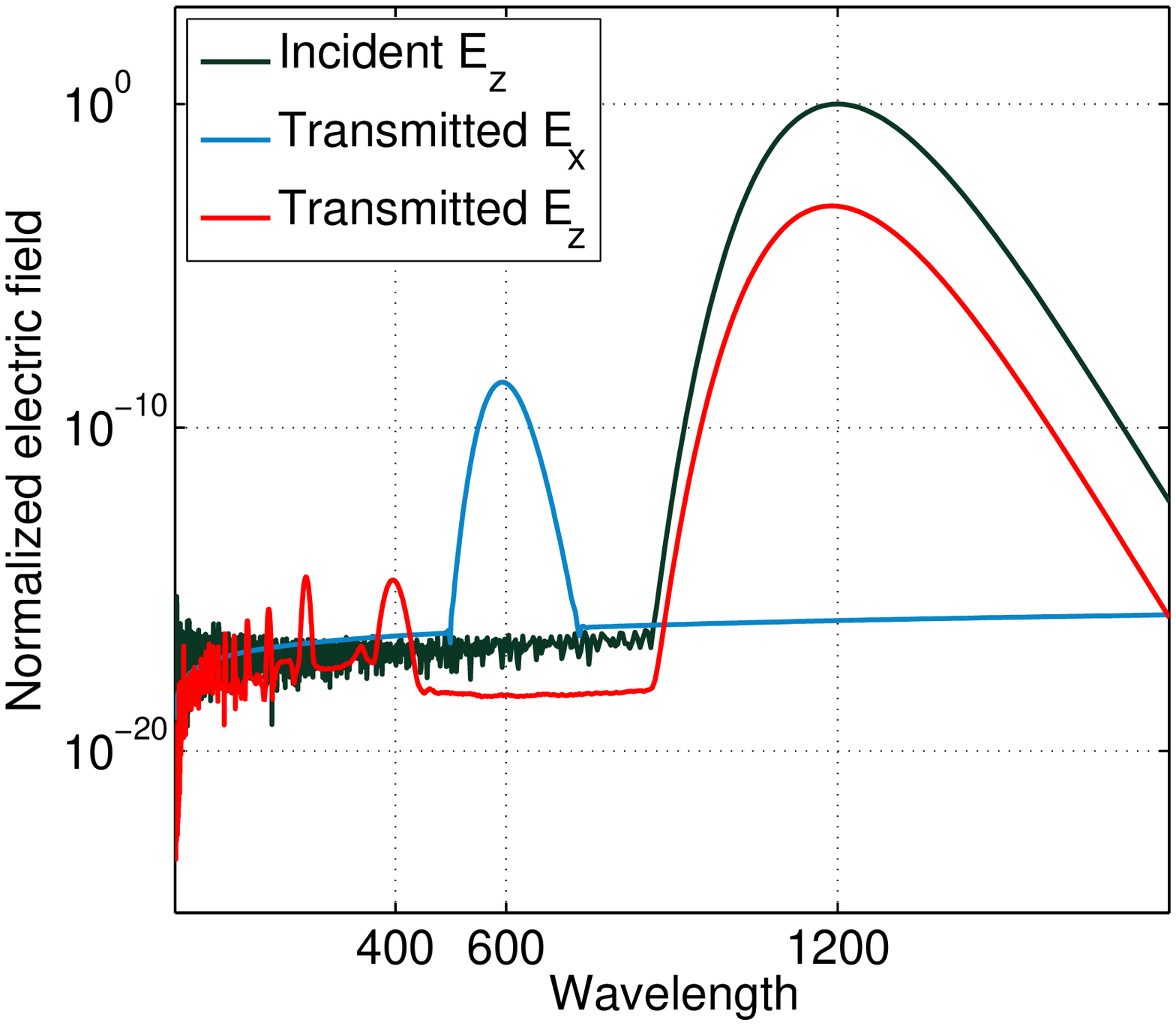}}
  {\includegraphics[width=7cm]{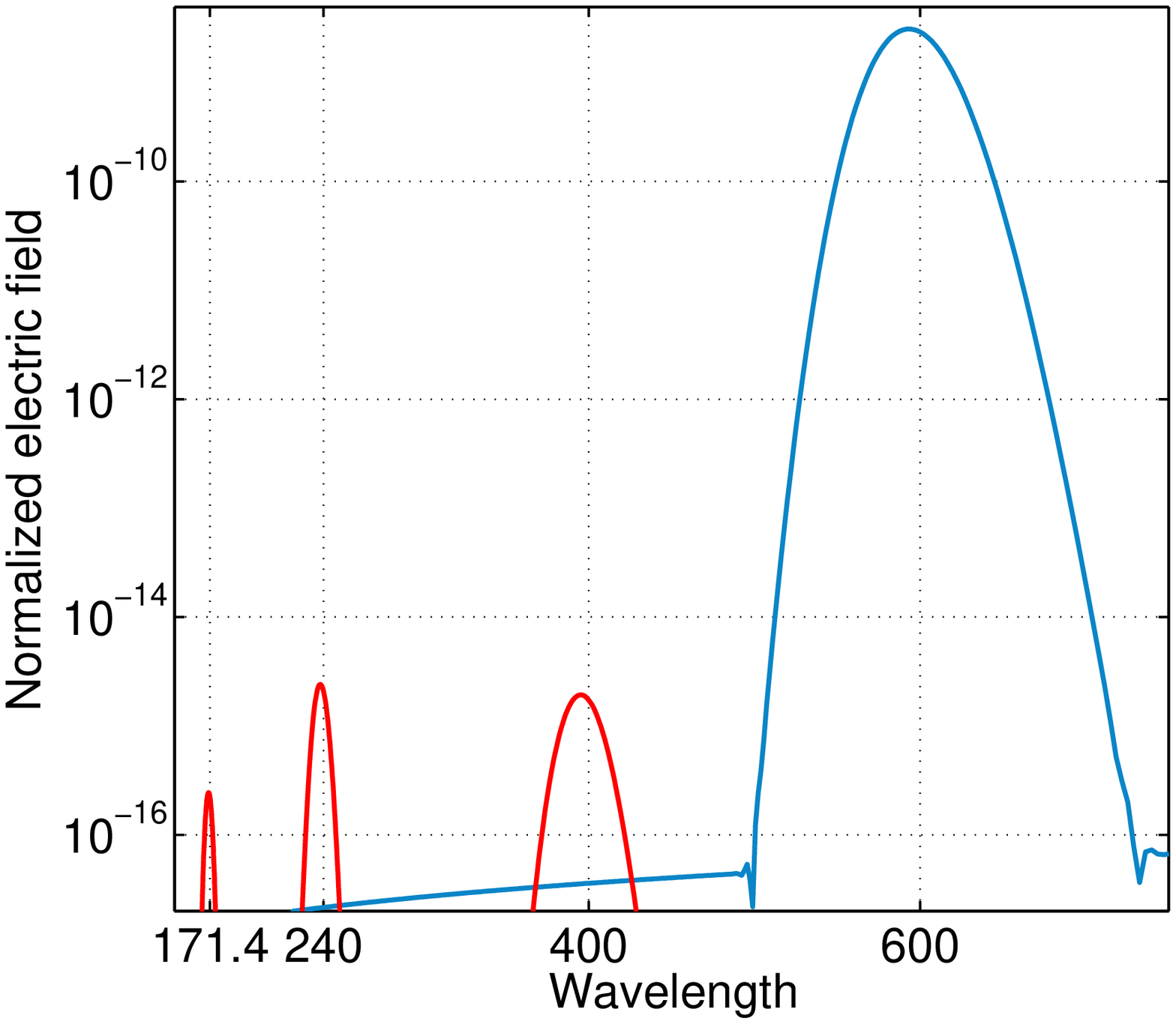}}
\caption{Spectrum with an L-shaped nanostructure. Left: Spectrum of incident wave $E_{z}$, transmitted wave $E_{x}, E_{z}$; right: Zoom-in plot of the spectrum.  }\label{L-shapedfrequency}
\end{figure}

Then, we take an investigation on the high order harmonic generation from the L-shaped metallic nanostructures (see Figure \ref{setup} (middle) for the setup). In Figure \ref{L-shapedtime}, we plot the time history of the linear response and the SHG.  In Figure \ref{L-shapedfrequency} (Left), spectrums of these responses are presented. Since the centrosymmetry is broken in this case,  the second order harmonic generation appears. As we can see from Figure \ref{L-shapedfrequency} that the fundamental wave ($1200$) can be frequency-doubled after propagating through the nanostructure, and  the higher order harmonic waves, such as the third-order harmonic generation ($400$), the fifth-order harmonic generation ($240$) and the seventh-order harmonic generation ($171.4$) in the transmitted $E_z$, are also captured in the simulation. The above results associated with both the rectangular and the L-shaped metallic nanostructures are in a good agreement with those presented in  \cite{Liu2010}.

\begin{figure}[H]
\centering
  {\includegraphics[height=6cm,width=7cm]{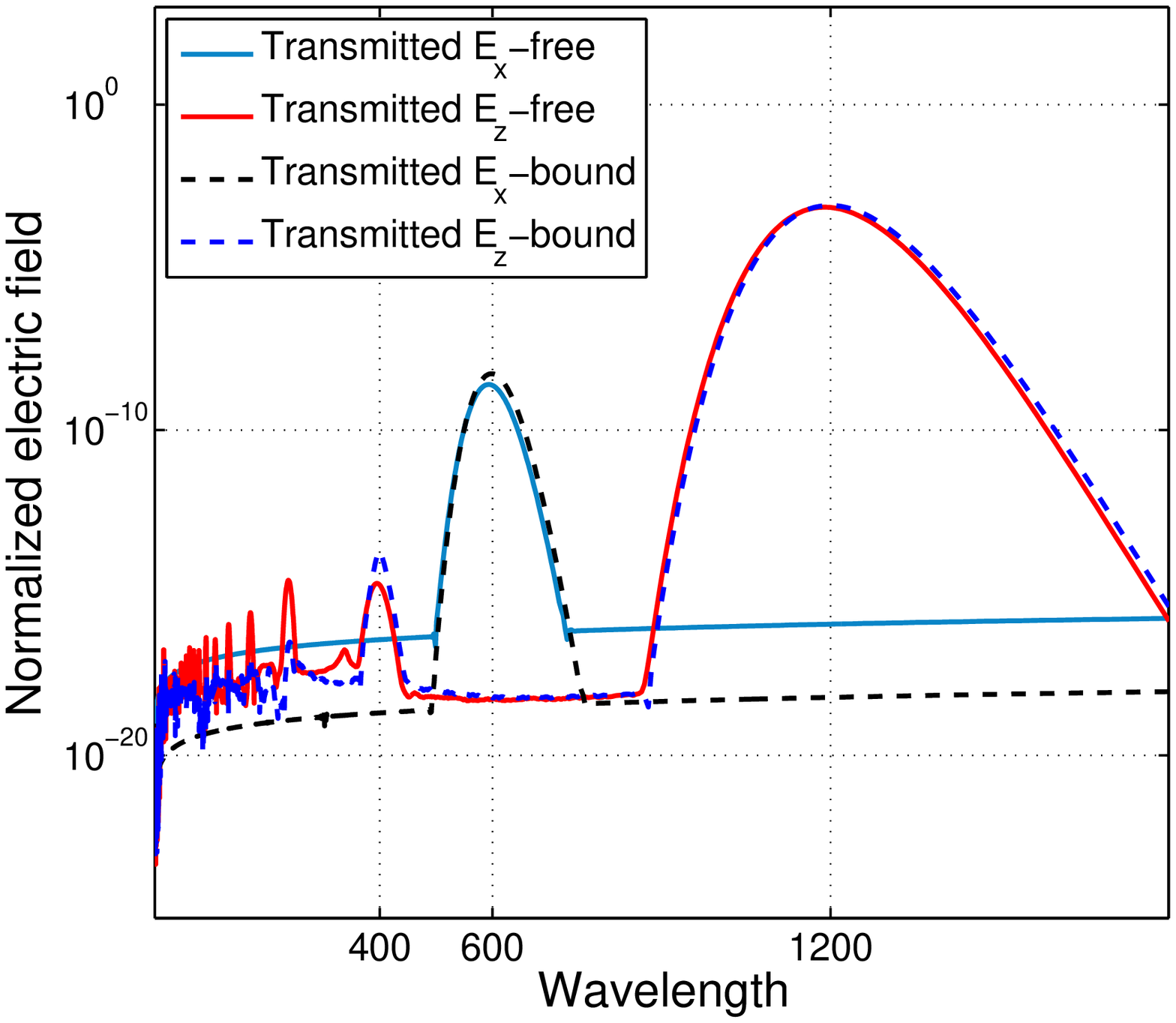}}
  {\includegraphics[height=6cm,width=7cm]{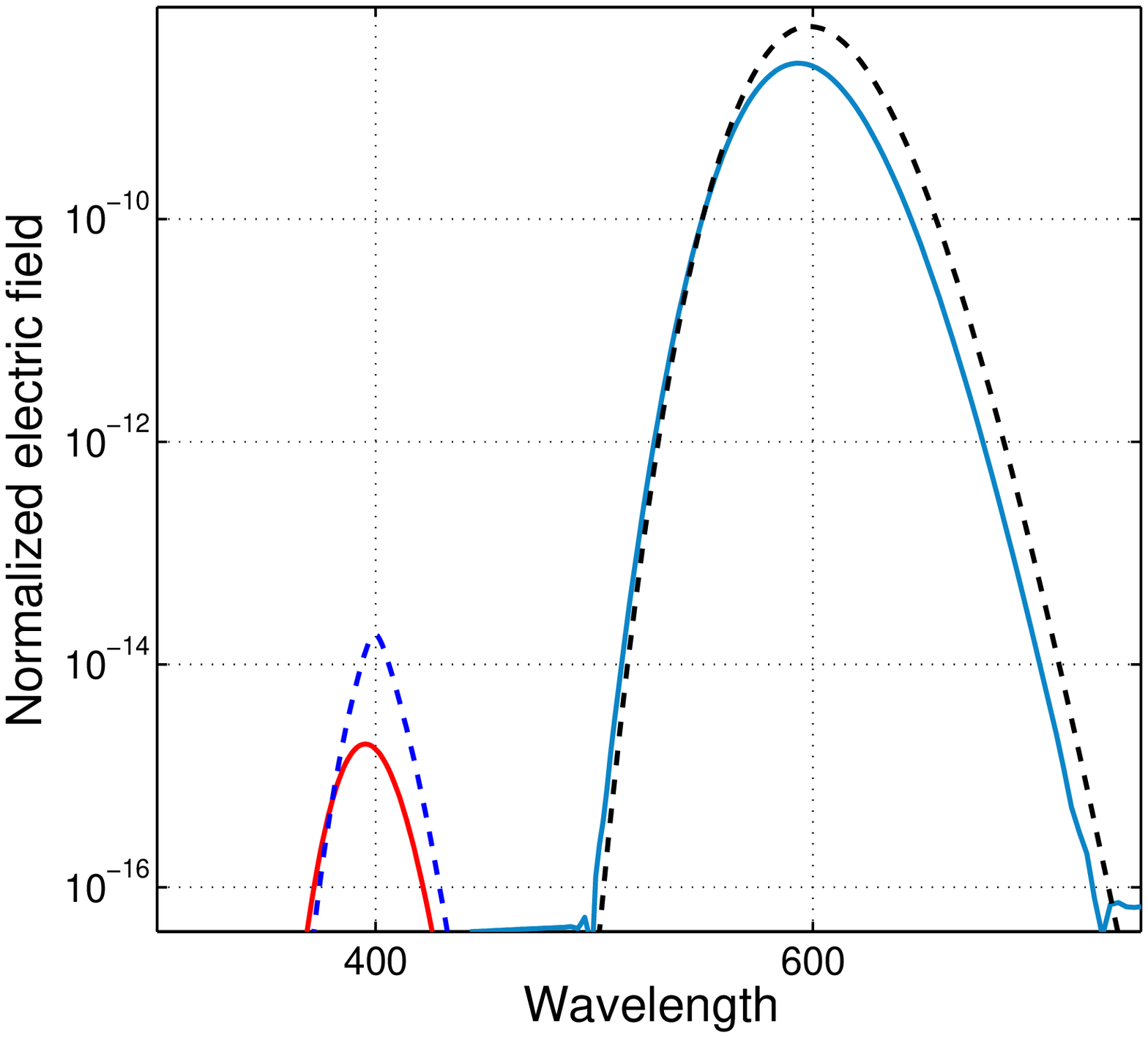}}
\caption{Spectrum from a nanostructure with bound electrons. Left: Spectrum of transmitted wave $E_{x}, E_{z}$; right: Zoom-in plot of the spectrum. }\label{L-shapedfrequency-bound}
\end{figure}

Finally, we investigate the high order harmonic generation from an L-shaped metallic nanostructure, with a rectangular non-metallic material occupied nanostructure being located at its corner (see the right of Figure \ref{setup}), to study the
influence of bound electrons on the generation of high order harmonic waves.
The simulation setting is the same as the L-shaped case.
In the simulations, we make a selection as ${n}_b=58.0$, ${\omega}_b=\sqrt{{n}_b {q}_e^2/{m}_e}$, ${\gamma}_b=0.8{\gamma}$.
In Figure \ref{L-shapedfrequency-bound} (Left), spectrums of the linear and nonlinear optical responses are presented, in comparison with spectrums obtained from the L-shaped metallic nanostructure  in the previous test.    It can be observed  that the high-order harmonics are enhanced when the interaction between the bound electrons and the external electromagnetic waves has been taken into account.

\subsection{Nonlocal effect}

For small optical particles, especially for particles with size down to subwavelength, apart from the surface plasmon, the bulk plasmon could be excited as well. In this test, we solve the MNHD model to investigate the nonlocal effect for the Ag nanowire. As in  \cite{Hiremath2012}, we choose the nanowire with radius ${r}=2.0$, the plasma frequency ${\omega}_p=2.885\times10^{-2}$, and the damping constant ${\gamma}=0.01{\omega}_p$. PML boundary conditions are employed in both $x-$ and $y-$directions. To resolve the ECS in the time domain, we introduce the $x-$polarized incident wave propagating in $y-$direction by the TF/SF technique and collect the Fourier-transformed total field and scattered field on the TF/SF boundary. The ECS is
calculated as follows
\begin{eqnarray*}\label{scs}
  C_{sca}(\omega) &=& \frac{\int_F\mathbf n\cdot \mathbf S_{sca}(\omega)dl}{|\mathbf S_{inc}(\omega)|}~, \\
  C_{abs}(\omega) &=& -\frac{\int_F\mathbf n\cdot \mathbf S_{tot}(\omega)dl}{|\mathbf S_{inc}(\omega)|}~, \\
  C_{ext}(\omega) &=& C_{sca}(\omega)+C_{abs}(\omega)~,
\end{eqnarray*}
where $\mathbf n$ denotes the outward unit normal to the TF/SF boundary $F$, and $\mathbf S_{\vartriangle}$ denotes the time averaged Poynting vector
\begin{equation}\label{energydensity}
  \mathbf S_{\vartriangle}(\omega)=\frac{1}{2}\mathbf E_{\vartriangle}(\omega)\times\mathbf H_{\vartriangle}^*(\omega)~.
\end{equation}

\begin{figure}[H]
  \centering
  \includegraphics[width=7cm]{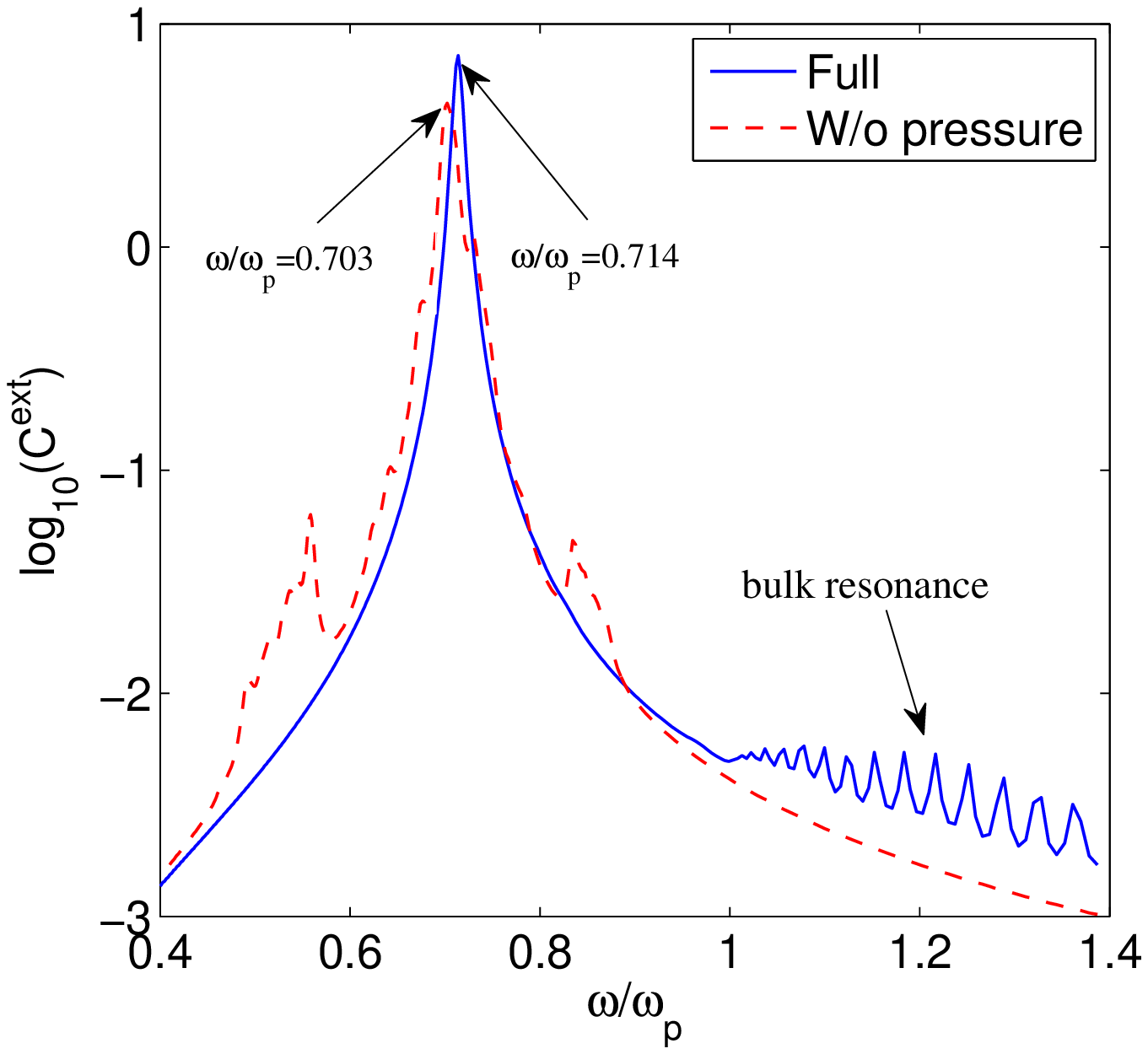}
  \includegraphics[width=7cm]{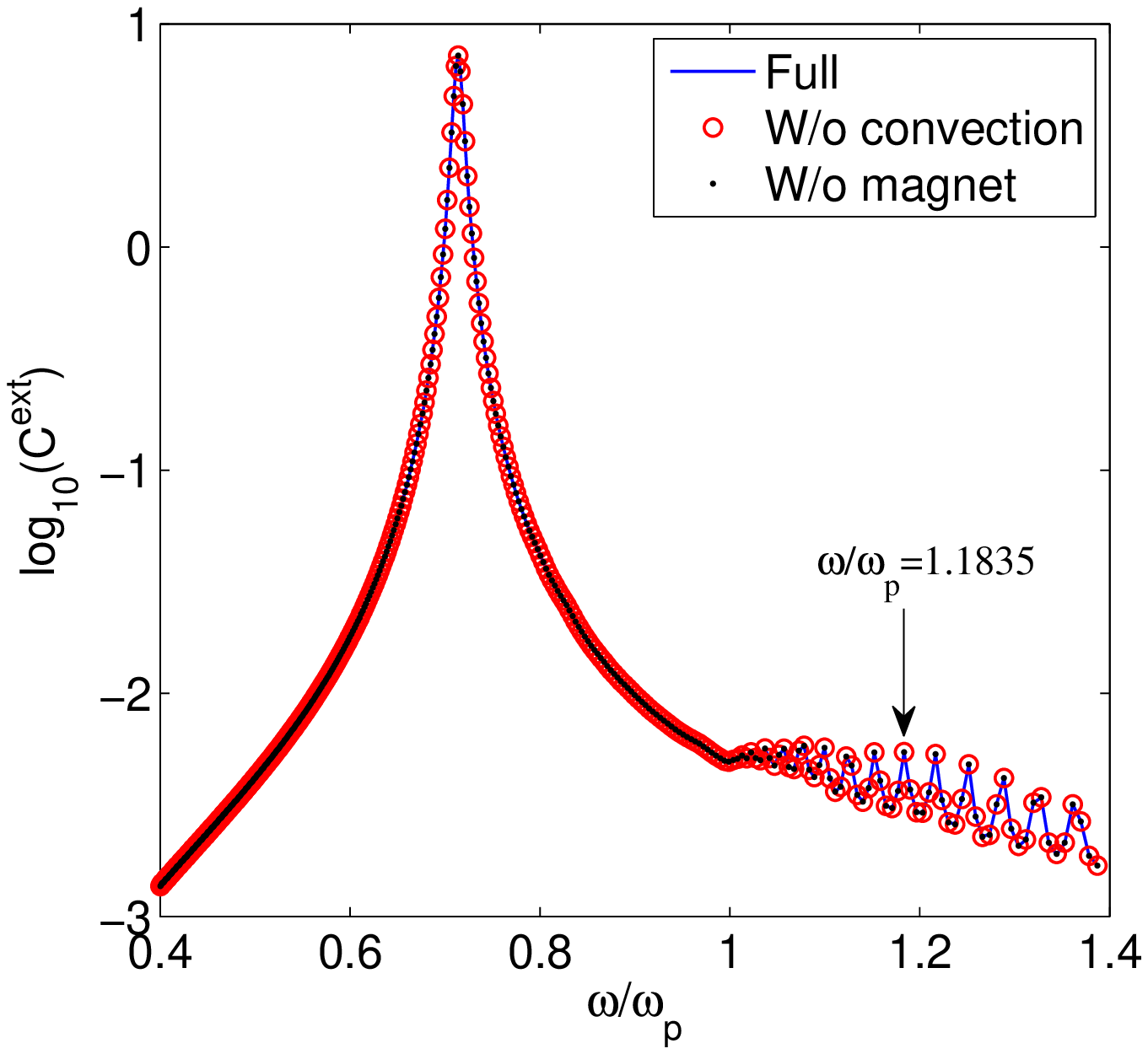}\\
  \caption{ECS calculated by nonlinear HD model for Ag nanowire with radius ${r}=2.0$.}\label{ecsnn}
\end{figure}

\begin{figure}[H]
  \centering
  \includegraphics[width=7cm]{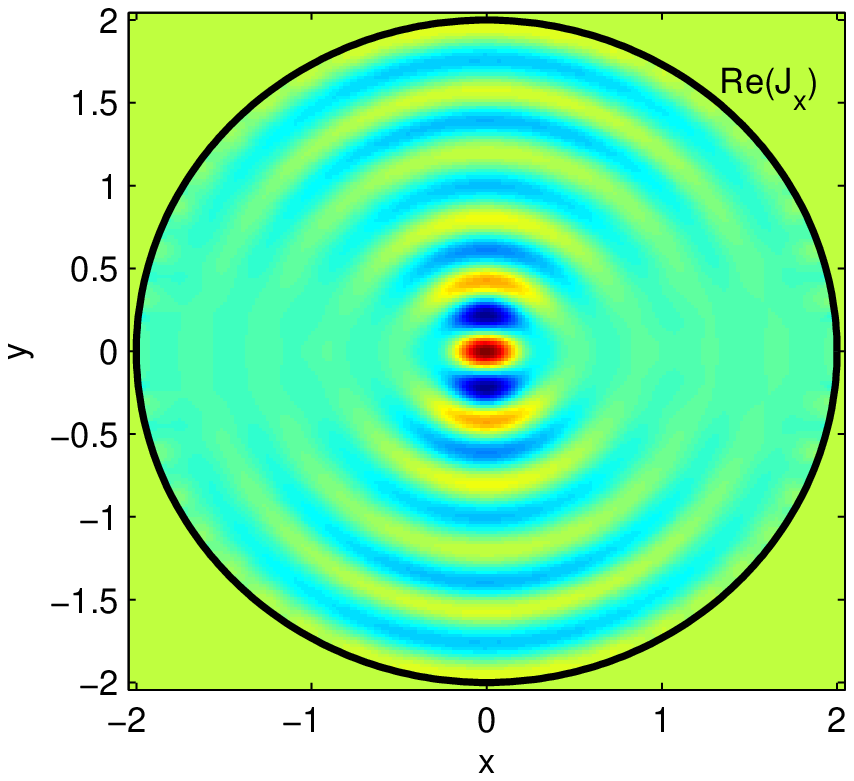}
  \includegraphics[width=7cm]{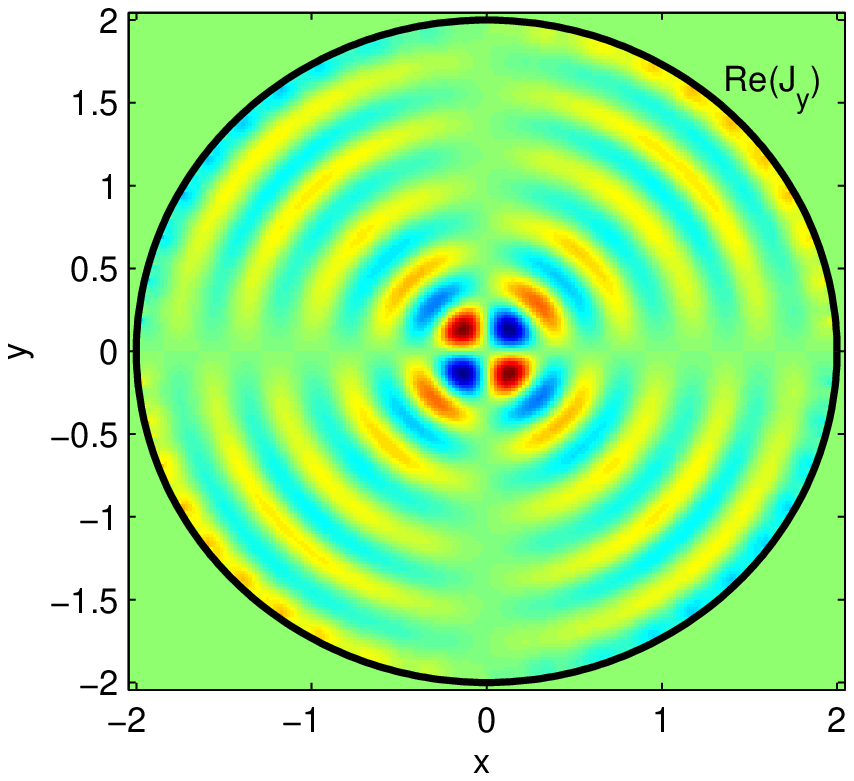}
  \caption{The bulk plasmon resonances of the current density ($\omega/\omega_p=1.1835$).}\label{reso}
\end{figure}

 We try to recover the normalized ECS from $0.4{\omega}_p$-$1.4{\omega}_p$ \cite{Hiremath2012}.  In the simulations, we adopt a short Gaussian pulse with ${t}_b=2.998,~ {\lambda}_0=300$ carrying effective information between the range $0.4{\omega}_p$-$1.4{\omega}_p$ to perform a broad-band calculation. We simulate the optical interaction for a long time such that the scattered field decays adequately into a steady state and we set the final time ${T}=5.996\times10^4$ for this purpose. In Figure \ref{ecsnn},
we present the ECS being normalized with respect to the diameter of nanowire as a function of the normalized angular frequency $\omega/\omega_p$.
It can be seen from Figure \ref{ecsnn} that  the bulk plasmon resonances beyond the plasma frequency can be excited with the MNHD model.  A slight blue-shift (from $\omega/\omega_p=0.703$ to $\omega/\omega_p=0.714$) of the surface plasmon resonance has been retrieved in the ECS as well. In order to make a further exploration on the source  for the appearance  of bulk plasmon resonances, we switch off the different nonlinear terms (quantum pressure term: $k\rho^{5/3}$, convection terms: $\rho u_iu_j, i,j=x,y,z$, magnetic terms: $u_iH_j-u_j-H_i, (i,j)=(x,y),(y,z),(z,x)$) in the equation \eqref{Eq:conservation2},  and make the calculation of the ECS, respectively. It can be observed from Figure \ref{ecsnn} that, apart from the ECS calculated from the MNHD model without the quantum pressure term, all other ECSs are in perfect match with the calculated ECSs associated to the full MNHD model. This indicates that, among all three terms,  the quantum pressure makes a unique contribution to   appearance  of  the nonlocal effect.
In Figure \ref{reso}, we provide with the Fourier-transformed current density at $\omega/\omega_p=1.1835$. As it is shown, the resonances,  namely the bulk plasmon, get excited in the bulk of nanowire.

\section{Conclusions}
In this paper, we develop a RKDG method for Maxwell equations nonlinearly coupled with  gas dynamic models with both the quantum pressure and bound electrons being taken into account. Proper initial and boundary conditions, coupled with the DG method,  have been designed for the efficient numerical simulation. Numerical results show that the high order harmonic waves can be produced from the L-shaped nanostructure and the rectangular nanostructure, and the bulk plasmon resonance can be excited for the metallic nanowire.
The effect of bound electrons on the generation of high order harmonic waves have been confirmed in numerics. Meanwhile, a switch-off comparison confirms that the quantum pressure term in the MNHD model is essentially important  for the production of nonlocal effects. Theoretical study on the quantum pressure terms,  numerical investigations on the influence of spill-out electrons on these important and complex optical phenomena, and three dimensional simulations will be envisioned  in our future work.

\section*{Acknowledgments}

The work of M. Lyu is partially supported by the Chongqing University Graduate Student Research Innovation Project (Project No. CYS15018). The work of L. Jiang is partially supported by the Research Grants Council of Hong Kong GRF (Grant No. 17209918). The work of M. Li is partially supported by NSFC (Grant Nos. 11871139, 11701055). The work of L. Xu is partially supported by a Key Project of the Major Research Plan of NSFC (Grant No. 91630205) and a grant of NSFC (Grant No. 11771068).


\begin{thebibliography}{00}

\bibitem{Anker2008} J. N. Anker, W. P. Hall, O. Lyandres, N. C. Shah, J. Zhao, R. P. Van Duyne, Biosensing with Plasmonic Nanosensors, Nat. Mater., 7 (2008) 442-453 .

\bibitem{Fengyan2017}V. A. Bokil, Y. Cheng, Y. Jiang, F. Li, Energy stable discontinuous Galerkin methods for Maxwell's equations in nonlinear optical media, J. Comput. Phys., 350 (2017) 420-452.


\bibitem{Bozhevolnyi2006} S. I. Bozhevolnyi, V. S. Volkov, E. Devaux, J.-Y. Laluet, T. W. Ebbesen, Channel plasmon subwavelength waveguide components including interferometers and ring resonators, Nature, 440 (2006) 508-511.


\bibitem{Chen2012} X. Chen, O. Nadiarynkh, S. Plotnikov, P. J. Campagnola, Second harmonic generation microscopy for quantitative analysis of collagen fibrillar structure, Nat. Protoc., 7 (2012) 654-669.

\bibitem{Ciraci2012} C. Ciraci, R.T. Hill, J.J. Mock, Y. Urzhumov, A.I. Fernandez-Dominquez, S.A. Maier, J.B. Pendry, A. Chilkoti, D.R. Smith, Probing the ultimate limits of plasmonic enhancement, Science, 337 (2012) 1072-1074.

\bibitem{Cockburn1998} B. Cockburn, C.-W. Shu, The Runge-Kutta discontinuous Galerkin method for conservation laws V: multidimensional systems, J. Comput. Phys., 141 (1998) 199-224.



\bibitem{Fang2016} M. Fang, Z. Huang, W. E. I. Sha, X. Xiong, X. Wu, Full hydrodynamic model of nonlinear electromagnetic response in metallic metamaterials, Prog. Electromagn. Res. 157 (2016) 63.




\bibitem{Gottlieb2001} S. Gottlieb, C.-W. Shu, E. Tadmor, Strong stability preserving high order time discretization methods, SIAM Review, 43 (2001) 89-112.

\bibitem{Hille2016} A. Hille, M. Moeferdt, C. Wolff, C. Matyssek, R. Rodr\'{\i}guez-Oliveros, C. Prohm, J. Niegemann, S. Grafstr\"{o}m, L.M. Eng, K. Busch, Second harmonic generation from metal nano-particle resonators: Numerical analysis on the basis of the hydrodynamic Drude model, J. Phys. Chem. C, 120 (2016) 1163-1169.

\bibitem{Hiremath2012} K. R. Hiremath, L. Zschiedrich, F. Schmidt, Numerical solution of nonlocal hydrodynamic Drude model for arbitrary shaped nano-plasmonic structures using N\'{e}d\'{e}lec finite elements, J. Comput. Phys., 231 (2012) 5890-5896.





\bibitem{Jackson1999} J. D. Jackson, Classical  Electrodynamics, 3rd ed., Wiley (1999).

\bibitem{Klein2006} M. W. Klein, C. Enkrich, M. Wegener, S. Linden, Second-harmonic generation from magnetic metamaterials, Science, 313 (2006) 502-504.

\bibitem{Klein2007} M. W. Klein, M. Wegener, N. Feth, S. Linden, Experiments on second- and third-harmonic generation from magnetic metamaterials, Opt. Express, 15 (2007) 5238-5247.

\bibitem{Krasavin2016} A. V. Krasavin, P. Ginzburg, G. A. Wurtz, A.V. Zayats, Nonlocality-driven supercontinuum white light generation in plasmonic nanostructures, Nat. Commum. 7 (2016) 11497.



\bibitem{Lal2008} S. Lal, S. E. Clare, N. J. Halas, Nanoshell-enabled photothermal cancer therapy: impending clinical impact, Acc. Chem. Res., 41 (2008) 1842-1851.

\bibitem{Liu2010} J. Liu, M. Brio, Y. Zeng, A. R. Zakharian, W. Hoyer, S. W. Koch, J. V. Moloney, Generalization of the FDTD algorithm for simulations of hydrodynamic nonlinear Drude model, J. Comput. Phys., 229 (2010) 5921-5932.

\bibitem{UPML} T. Lu, P. Zhang, W. Cai, Discontinuous Galerkin methods for dispersive and lossy Maxwell¡¯s equations and PML boundary conditions, J. Comput. Phys., 200 (2004) 549-580.

\bibitem{Mortensen2014} N.A. Mortensen, S. Raza, M. Wubs, T. Sondergaard, S.I. Bozhevolnyi, A generalized non-local optical response theory for plasmonic nanostrutures, Nat. Commun., 5 (2014) 3809.

\bibitem{Niesler2009} F. B. P. Niesler, N. Feth, S. Linden, J. Niegemann, J. Gieseler, K. Busch, M. Wegener, Second-harmonic generation from split-ring resonators on a GaAs substrate, Opt. Lett., 34 (2009) 1997-1999.

\bibitem{Parr1994} R. G. Parr, W. Yang, Density-Functional Theory of Atoms and Molecules, Oxford Univ. Press, 1994.

\bibitem{Patel1965} C. K. N. Patel, Efficient Phase-Matched Harmonic Generation in Tellurium with a $CO_2$ Laser at $10.6 \mu$, Phys. Rev. Lett., 15 (1965) 1027-1030.

\bibitem{Scalora2010} M. Scalora, M. A. Vincenti, D. de Ceglia, V. Roppo, M. Centini, N. Akozbek, M. J. Bloemer, M. Centini, Second and Third Harmonic Generation in Metal-Based Nanostructures, Phys. Rev. A, 82 (4) (2010) 5929-5937.

\bibitem{Schmitt2016} N. Schmitt, C. Scheid, S. Lanteri, A. Moreau, J. Viquerat, A DGTD method for the numerical modeling of the interaction of light with nanometer scale metallic structures taking into account non-local dispersion effects, J. Comput. Phys., 316 (2016) 396-415.

\bibitem{Sonnenberg1968} H. Sonnenberg, H. Heffner, Experimental Study of optical second-harmonic generation in silver, J. Opt. Soc. Am., 58 (1968) 209-212.

\bibitem{Taflove1995}A. Taflove, Computational Electrodynamics: The Finite-Difference Time-Domain Method, Artech House, London, 1995.

\bibitem{Toscano2015} G. Toscano, J. Straubel, A. Kwiatkowski, C. Rockstuhl, F. Evers, H. Xu, N. A. Mortensen, M. Wubs, Resonance shifts and spill-out effects in self-consistent hydrodynamic nanoplasmonics, Nat. Commun., 6 (2015) 7312.

\bibitem{Vidal2018} F. Vidal-Codina, N. C. Nguyen, S.-H. Oh,  J. Peraire, A hybridizable discontinuous Galerkin method for computing nonlocal electromagnetic effects in three-dimensional metallic nanostructures, J. Comput. Phys., 355 (2018) 548-565.



\bibitem{Wasylczyk2005} P. Wasylczyk, I. A. Walmsley, W. Wasilewski, C. Radzewicz, Broadband noncollinear optical parametric amplifier using a single crystal, Opt. Lett., 30 (2005) 1704-1706.

\bibitem{Zeng2009} Y. Zeng, W. Hoyer, J. Liu, S.W. Koch, J.V. Moloney, A classical theory for second-harmonic generation from metallic nanoparticles, Phys. Rev. B, 79 (2009) 235109.






\end{thebibliography}
\end{document}